\documentclass[aps,prd,showpacs,floatfix,a4paper,notitlepage]{revtex4-1}
\usepackage{amsfonts,amsmath,units,wasysym,epsfig,graphicx,verbatim,color,subfigure,graphicx,bm,mathrsfs,lipsum,hyperref}
\usepackage[utf8]{inputenc}
\usepackage[normalem]{ulem}  

\newcommand{\caen}{\affiliation{Normandie Univ, ENSICAEN, UNICAEN, CNRS/IN2P3, LPC Caen, 14000 Caen, France}}
\newcommand{\parisCit}{\affiliation{Laboratoire Univers et Th\'eories, CNRS, Observatoire de Paris, Universit\'e PSL, Universit\'e Paris Cit\'e, 5 place Jules Janssen, 92195 Meudon, France}}
\newcommand{\liege}{\affiliation{Space Sciences, Technologies and Astrophysics Research (STAR) Institute, Universit\'e de Li\`ege, B\^at. B5a, 4000 Li\`ege, Belgium}}
\newcommand{\brussels}{\affiliation{Institut d’Astronomie et d’Astrophysique, Universit\'e Libre de Bruxelles, CP 226, B-1050 Brussels, Belgium}}
\begin{document}
\title{Generalised description of Neutron Star matter with nucleonic Relativistic Density Functional}
\author{  P. Char} \email{prasanta.char@uliege.be } \liege
  \author{C. Mondal} \email{chiranjib.mondal@ulb.be} \caen \brussels
\author{F. Gulminelli} \email{gulminelli@lpccaen.in2p3.fr} \caen
    \author{M. Oertel} \email{micaela.oertel@obspm.fr} \parisCit

\begin{abstract}
 In this work, we propose a meta-modelling technique to nuclear matter on the basis of a relativistic density functional with density-dependent couplings. Identical density dependence for the couplings both in the isoscalar and isovector sectors is employed. We vary the coupling parameters of the model to capture the uncertainties of the empirical nuclear matter parameters at saturation. Then, we construct a large ensemble of unified equations of state in a consistent manner both for clusterized and uniform matter in $\beta$-equilibrium at zero temperature. Finally, we calculate neutron star properties to check the consistency with astrophysical observations within a Bayesian framework. Out of the different sets of astrophysical data employed, constraint on tidal deformability from the GW170817 event was found to be the most stringent in the posteriors of different neutron star properties explored in the present study.  We demonstrate in detail the impact of the isovector incompressibility ($K_{sym}$) on high-density matter that leads to a considerable variation in the composition of neutron star matter. A couple of selected models with extreme values of $K_{sym}$, which satisfy various modern nuclear physics and neutron star astrophysics constraints, are uploaded in the \textsc{CompOSE} \cite{Typel:2013rza} database for use by the community. 
 \end{abstract}

\maketitle
\section{Introduction}
The theoretical description of strongly interacting systems of baryons and mesons leads to the construction of the equation of state (EOS) of dense matter which becomes the foundational basis for the calculation of the properties of infinite nuclear matter and 
neutron stars. Despite the tremendous advancements in theoretical and experimental nuclear physics during the last decades, our knowledge of the EOS is still incomplete. Quantum Chromodynamics, although being the fundamental theory of strong interaction, does not provide any useful solution at the densities relevant to the finite and infinite nuclear systems \cite{Glendenning:1997wn}. Hence, there have been numerous attempts to construct an effective theory of matter relevant for the hadronic degrees of freedom. These approaches can be broadly divided into two categories: (1) ab-initio many body methods, and (2) phenomenological methods \cite{Oertel:2016bki}. In recent years, ab-initio approaches have seen many developments and have in particular benefited from the construction of new nuclear interaction potentials within chiral effective field theory ($\chi$-EFT)~\cite{Drischler:2015eba,Drischler:2021kxf}. On the other hand, phenomenological models have been quite useful due to their simpler structures and reduced computational costs which allow us to compute the EOS for the full range of temperatures, baryon number densities and compositions, needed to describe the objects of interest in compact star astrophysics, i.e. (proto-)neutron stars, core-collapse supernovae (CCSN) and binary neutron star (BNS) mergers.  Traditionally, the parameters of the phenomenological interactions are optimized by fitting a variety of experimental nuclear data. These approaches are usually expressed by the energy density functionals, either as non-relativistic ones with zero range (Skyrme type) or finite range interactions (Gogny type) \cite{Dutra:2012mb}, or as (special) relativistic ones (relativistic mean field type) \cite{Dutra:2014qga}. All of them have been widely used in astrophysical applications such as to describe the composition and the thermodynamic properties of (proto-)neutron star matter, CCSN, and BNS mergers. Relativistic models are better suited to describe high-density matter as they have built-in causality, and the inclusion of different non-nucleonic degrees of freedom can be handled more coherently. They originated as a field theory expressed by a Lagrangian density where hadrons interact via the exchange of scalar and vector mesons representing the attractive and repulsive component, respectively, of the nuclear force to describe the saturation properties of symmetric nuclear matter \cite{Walecka:1974qa,Serot:1984ey}. Later, various extensions were implemented including an isovector interaction, see e.g.~\cite{Mueller:1996pm,Todd-Rutel:2005yzo}, non-linear self-interaction of mesons, see e.g ~\cite{Reinhard:1986qq,Rufa:1988zz,Sugahara:1993wz}, to describe the in-medium effects and characterize the high-density behavior within the theory. A different class of relativistic density functional models was also introduced in the form of density dependent couplings, see Refs.~\cite{Fuchs:1995as,Typel:1999yq} for the first realisations.

Various properties for the ground and excited states of different nuclei, \textit{e.g.} binding energy, charge radii, isobaric analogous states, giant resonances or data from heavy ion collisions are used to optimize these phenomenological interactions \cite{Klupfel:2008af,Dobaczewski:2014jga} and very satisfactory results have been obtained. Recent major developments of these models concern mainly the isovector sector of the nuclear interaction \cite{Horowitz:2000xj,Todd-Rutel:2005yzo} in relation with experimental progress \cite{HADES:2020lob,Tanaka:2021oll}. Particularly, recent laboratory experiments like PREX-II and CREX based on the parity violating electron scattering have measured the neutron skin thickness of $^{208}$Pb and $^{48}$Ca, respectively, albeit with relatively large uncertainties~\cite{Abrahamyan:2012gp,PREX:2021umo,CREX:2022kgg}. {The impact of these results are investigated in several theoretical studies in the context of underlying correlations involving properties of finite nuclei \cite{Reinhard:2022inh, Yuksel:2022umn, Mondal:2021jlo} and neutron stars, in particular, the radius and tidal deformability \cite{Mondal:2022cva}.} A future more precise measurement would clearly be very important to pin down the symmetry energy part of the nuclear interaction even further \cite{Becker:2018ggl}.  

Multi-messenger observations of neutron stars (NSs) provide excellent probes of supranuclear matter to complement the knowledge gained from terrestrial experiments. NSs are gravitationally bound objects, and the solutions of the general relativistic structure equations of NSs allow to obtain a sequence of mass and radius which have a 
direct correspondence to the EOS of NS matter~\cite{Lattimer:2000nx,Lattimer:2006xb,Lattimer:2012nd}. Recent important constraints from NS observations came with the observations of massive pulsars ($\gtrsim 2 M_\odot$) with precisely determined masses that ruled out many EOS models \cite{Antoniadis:2013pzd,Fonseca:2021wxt}. The detection of gravitational waves from the BNS merger event GW170817 by the LIGO-Virgo collaboration with a first value for the NS tidal deformability together with the corresponding electromagnetic counterparts ushered a new era of the exploration of matter under extreme conditions \cite{TheLIGOScientific:2017qsa,LIGOScientific:2017ync}. Parallel improvements in the X-ray observations with the NICER instrument onboard the ISS has provided two simultaneous mass-radius measurements of neutrons stars \cite{Riley:2019yda,Miller:2019cac,Riley:2021pdl,Miller:2021qha}. In order to analyse these astrophysical measurements in terms of the NS EOS, agnostic versions of the high-density EOS in the form of piecewise polytropes \cite{Read:2008iy}, spectral parameterization \cite{Lindblom:2010bb,Lindblom:2012zi,Lindblom:2013kra}, or non-parametric Gaussian process-based sampling \cite{Landry:2018prl,Essick:2019ldf,Landry:2020vaw} have been put forward. Although these approaches are free from any assumptions on any particular nuclear interaction model, they only allow to pin down the EOS of $\beta$-equilibrated matter without any understanding of the underlying matter composition and interactions. 

To bridge the gap between the microscopic theory of dense matter and the required agnosticity, a nuclear meta-modelling approach EOS has recently been developed based on a Taylor expansion, truncated at fourth order, of the energy per baryon in density and asymmetry around saturation~\cite{Margueron:2017eqc, Margueron:2017lup}. In this technique, the non-trivial dependence on Fermi momentum at low density was achieved by separating the kinetic and potential parts of the functional. The expansion coefficients of the potential part are expressed as functions of empirical nuclear matter parameters (NMPs) which are constrained by nuclear experiments for the two lowest orders. 
These latter include the binding energy per nucleon $E_{sat}$, the saturation density $n_{sat}$, incompressibilty  $K_{sat}$, the symmetry energy at saturation $E_{sym}$, the slope of symmetry energy $L_{sym}$, isovector incompressibility $K_{sym}$. A purely nucleonic composition is hypothesized for the baryonic component. 
This approach called  ``nucleonic meta-modelling" incorporates diverse possible predictions for the dense matter EOS and composition through the experimental uncertainties assumed for the NMPs.  It  provides a generic EOS model that encompasses nuclear physics knowledge from laboratory experiments and astrophysical observations within a Bayesian framework. {The model was successfully applied in the recent years to obtain controlled predictions of different astrophysical quantities \cite{Thi:2021jhz,Thi:2021hai,Mondal:2022cva,Mondal:2023gbf}.}

The main limitation of this approach to the nucleonic meta-modelling is that it uses a non-relativistic formulation and that causality has to be imposed by hand a posteriori~\cite{Margueron:2017eqc}. In addition, it is not straightforward to include other non-nucleonic degrees of freedom such as hyperons or mesons consistently in the formulation. Only an extension with a phase transition to quark matter on the basis of a Maxwell/Gibbs construction between a nucleonic model on one side and a quark model on the other side is feasible~\cite{Mondal:2023gbf,Pfaff:2021kse,2023arXiv230318133G}. In order to improve these aspects, we introduce here a relativistic meta-modelling. For the present work we will limit ourselves to a purely nucleonic composition of high-density baryonic matter, but the formulation allows for an extension to other degrees of freedom. We have chosen a relativistic energy density functional with density dependent couplings for this purpose. This type of model was first introduced by Fuchs and Lenske~\cite{Fuchs:1995as,Lenske:1995wyj} for the description of finite nuclei. The functional form of such density dependence was discussed in detail by Typel and Wolter \cite{Typel:1999yq}. Subsequently, the model was applied to study hypernuclei and neutron star matter as well \cite{Keil:1999hk,Hofmann:2000vz,Hofmann:2000mc}. For an in-depth review of the relativistic energy density functionals with density dependent couplings see \textcite{Typel:2018cap} and references therein. In recent years, the choice of parameters for these models has been optimized with Bayesian techniques which allows to include consistently the different nuclear physics and astrophysics constraints into nuclear modelling~\cite{Traversi:2020aaa,Malik:2022zol,Zhu:2022ibs,Beznogov:2022rri,Malik:2023mnx,Salinas:2023nci,Huang:2023grj}. In contrast, in this work, we do not wish to find a best-fit model with all the available data, but we would rather like to analyze the parameter space of NMPs thoroughly and find a Lagrangian formalism for the meta-modelling that provides enough freedom to incorporate the uncertainties from the nuclear experiments and to predict the range of allowed NS EoS. 

We study the composition of the high-density matter within the relativistic meta-model, too. The composition of the NS core affects its properties significantly. The knowledge of the proton fraction is essential to our understanding of various astrophysical phenomena such as the cooling of NSs and nucleosynthesis \cite{Beloin:2018fyp} and kilonova properties of a BNS merger event, see \cite{Lunney:2020gjj} and references therein.  We are interested in exploring the full extent of the freedom in the isospin sector and thus for the proton fraction in NS matter. In particular, one of our objectives is to study the effect of the symmetry energy compressibility parameter $K_{sym}$, that is presently largely unconstrained by nuclear phenomenology. It has been shown very recently that the behavior of $K_{sym}$ may hold the key to explain the latest PREX and CREX results together \cite{Reed:2023cap}. We therefore give a special attention to explore the range of $K_{sym}$ and its correlations with other parameters within the framework of a relativistic meta-modelling approach.

The paper is organized as follows. In section \ref{formalism}, we discuss the construction of the relativistic meta-model and establish the motivation for a large exploration of the parameter space. In section \ref{data}, we outline the constraints used in this work. Then, we present our results and compare with previous works in the literature in section \ref{results}. Finally, we summarize our findings in section \ref{conclusion}.

\section{Relativistic Formulation of the meta-modelling}
\label{formalism}
In this section, we describe our implementation of the relativistic meta-modelling. We start with the definition of the NMPs and make a connection with the parameters of the energy density functional. 
Expanding the energy per nucleon of asymmetric nuclear matter $e(n_B,\delta)$ at baryon number density $n_B=(n_n+n_p)$, with $n_{n(p)}$ being the neutron (proton) density, and asymmetry $\delta=\frac{n_n - n_p}{n_B}$  
in a Taylor series around the equilibrium density of symmetric nuclear matter $n_{sat}$, the energy of symmetric matter $e_0(n_B)=e(n_B,\delta=0)$ reads, 

\begin{equation}
e_{0}(n_B) =  E_{sat}  + \frac{1}{2!} K_{sat} \left(\frac{n_B-n_{sat}}{3n_{sat}}\right)^2 + \frac{1}{3!}Q_{sat} \left(\frac{n_B-n_{sat}}{3n_{sat}}\right)^3 +\frac{1}{4!} Z_{sat} \left(\frac{n_B-n_{sat}}{3n_{sat}}\right)^4 + ...
\label{eq:SNM}
\end{equation}
The coefficients of the above expansion up to order four define the isoscalar parameters which are connected to the derivatives of $e_{0}$ with respect to baryon number density at saturation.
To characterize the behavior with respect to the asymmetry of nuclear matter, the so-called symmetry energy is defined as 

\begin{equation}
e_{sym}(n_B) = \frac{1}{2}\left.\frac{\partial^2 e(n_B,\delta)}{\partial \delta^2}‌\right|_{\delta=0} .
\end{equation}

Performing a Taylor expansion of $e_{sym}$ around saturation in a similar way as above allows to introduce the isovector NMPs as
\begin{equation}
e_{sym}(n_B) = E_{sym}  + L_{sym} \left(\frac{n_B-n_{sat}}{3n_{sat}}\right) + \frac{1}{2!} K_{sym} \left(\frac{n_B-n_{sat}}{3n_{sat}}\right)^2 + \frac{1}{3!}Q_{sym} \left(\frac{n_B-n_{sat}}{3n_{sat}}\right)^3 +\frac{1}{4!} Z_{sym} \left(\frac{n_B-n_{sat}}{3n_{sat}}\right)^4 + ... \label{eq:Symm}
\end{equation}
In symmetric nuclear matter (SNM), see Eq.~\eqref{eq:SNM}, $E_{sat}$ and $K_{sat}$ represent the binding energy per nucleon and incompressibility at saturation, respectively. In Eq.~\eqref{eq:Symm}, $E_{sym}, L_{sym}$ and $K_{sym}$  denote symmetry energy, slope and symmetry incompressibility at saturation, respectively. In these expansions, the third and fourth-order terms are the (symmetry) skewness ($Q_{sat}, Q_{sym}$) and kurtosis ($Z_{sat}, Z_{sym}$) respectively. 
Eqs.(\ref{eq:SNM}) and (\ref{eq:Symm}) give the definition of the NMPs, as they can be calculated from any arbitrary model with nucleonic degrees of freedom. For their sampling,  a self-consistent determination is performed in this paper based on
a meson exchange Langrangian density with density dependent couplings:
\begin{eqnarray}
\mathcal{L}_{\rm DD} &=& \overline{\psi}(i\gamma^\mu\partial_\mu - M)\psi 
+ \Gamma_\sigma(n_B)\sigma\overline{\psi}\psi 
- \Gamma_\omega(n_B)\overline{\psi}\gamma^\mu\omega_\mu\psi 
-\frac{\Gamma_\rho(n_B)}{2}\overline{\psi}\gamma^\mu{\boldsymbol \rho}_\mu \cdot {\boldsymbol \tau}
\psi  \nonumber \\
&+& \frac{1}{2}(\partial^\mu \sigma \partial_\mu \sigma - m^2_\sigma\sigma^2)
- \frac{1}{4}F^{\mu\nu}F_{\mu\nu} + \frac{1}{2}m^2_\omega\omega_\mu\omega^\mu 
-\frac{1}{4}\vec{B}^{\mu\nu}\vec{B}_{\mu\nu}+\frac{1}{2}m^2_\rho
{\boldsymbol \rho}_\mu \cdot {\boldsymbol \rho}^\mu .
\label{dldd}
\end{eqnarray}
Here, $\sigma,\,\omega_{\mu}$,and $\boldsymbol{\rho_{\mu}}$ are effective meson fields 
mediating the strong interaction among the nucleons, represented by the field $\psi$. We do not include the $\delta$ meson, i.e. a scalar isovector channel, as it does not have significant impact on the description of bulk properties of nuclear matter \cite{2004PhRvC..70e8801M}, although in recent times the impact to $\delta$ meson to mitigate apparent tension between the inferences from GW170817 and PREX-II results were pointed out in Ref. \cite{Li:2022okx}. To insure a maximum flexibility to the density dependence of the equation of state, only linear terms in the meson fields are considered in the Lagrangian 
while all the meson couplings $\Gamma_\sigma$, $\Gamma_\omega$, and $\Gamma_\rho$ are density dependent.   The functional form of the density dependence is chosen as  \cite{Gogelein:2007qa}
\begin{equation}
 \Gamma_i(n_B)=a_i+(b_i+d_i\,x^3)e^{-c_i\,x},
\label{GDFM}
\end{equation}
with $i=\sigma,\omega,\rho$, and $x = n_B/n_0$. $n_0$ is a normalisation density in general chosen close to, but not necessarily coincident with the saturation density.  We will denote our Lagrangian-based meta-model as GDFM model hereafter, after the initials of the authors that have first proposed this functional for for the density dependence of the couplings. Note that in contrast to the original paper  \textcite{Gogelein:2007qa} we do not include a correction term for the isoscalar vector $\omega$ channel close to saturation and we assume a priori all four parameters $a_i,b_i,c_i,d_i$ nonzero in all channels. Our choice of the particular form Eq.(\ref{GDFM}) is motivated by a few reasons. First of all, it puts all the couplings on the same footing for the density dependence; second, the isovector coupling has enough freedom to explore the uncertainties of the NMPs of interest. Of course, one can choose other forms depending on the requirement of the analysis. In a future work we will check to which extent our results depend on the particular choice for the functional form of the density dependence.
All in all the present model has 12 independent parameters determining the strength of the couplings and their density dependence. In our calculations, we have chosen to vary the dimensionless parameters $a_i,b_i, c_i, d_i$ fixing the meson masses and the scaling density $n_0$. For the latter, following \textcite{Gogelein:2007qa}, the following values have been assumed: $m_\omega=782.6$ MeV, $m_\rho=769$ MeV,  and $m_\sigma = 550$ MeV.  The nucleon mass is fixed to $M =938.9$ MeV. For the parameter $n_0$ we have taken the value $n_0 = 0.16$ fm$^{-3}$.  We remind the reader here that $n_0$ is not the saturation density. The latter is determined for a given set of model parameters by the condition of vanishing pressure in symmetric matter. 

To build an EOS for cold $\beta$-equilibrated matter which allows to calculate the structure of a NS, we need as well a description of the low-density crust region where nuclear clusters are formed. A consistent modelling of matter for the crust and the core has been found to be  important for the correct determination of NS radii and tidal deformabilities~\cite{Fortin:2014mya, Perot:2020gux, Suleiman:2021hre}. Here  we will  use a  unified approach, calculating   a consistent crust from the compressible liquid drop model (CLDM) approach of \textcite{Carreau:2019zdy}. This approach calculates the crust given a set of NMPs using the Taylor expansions of Eqs.(\ref{eq:SNM}) and(\ref{eq:Symm}). In our case these NMPs are calculated for each realization of the GDFM model. {To avoid any fictitious correlations imparted by high density and low density data on the NMPs we treat the third and fourth order NMPs below and above the saturation independently as described in Ref. \cite{Mondal:2022cva}.} 
The use of independent third and fourth-order parameters in principle breaks the unified character of the EoS; however, the effect of those parameters at low density is found to be negligible on the astrophysical quantities of interest for this paper.
Hence, in this way we obtain a unified EOS for each set of GDFM model parameters combining core and crust using the same set of first and second order NMPs.

\begin{table}
    \centering
    \begin{tabular}{c|c}
         \hline
         \hline 
         Parameter& Assumed range of values  \\
         \hline
         \hline
         $ n_{sat}$ (fm$^{-3}$)  & 
         $0.135,0.195$\\
         $ E_{sat}$ (MeV) & 
         $-17,-14$ \\
         $ K_{sat}$ (MeV) & 
         $150,350$ \\
         $ E_{sym}$ (MeV) & 
         $20,45$ \\
         $ L_{sym}$ (MeV) & 
         $20,180$ \\
         \hline
         \hline
    \end{tabular}
    \caption{Ranges of values considered for various nuclear empirical parameters}
    \label{tab:prior}
\end{table}
\section{Bayesian Analysis with nuclear and astrophysics constraints}
\label{data}
We perform a Bayesian study of the relativistic meta-model incorporating different types of constraints as filters for our generated samples of model parameters. The aim is then to predict global NS properties respecting all available constraints.
To that end, the posterior distributions of our set of model parameters $(\mathbf{X}= a_i, b_i, c_i, d_i)$ satisfying all the constraints $\mathbf{C}$ are defined as,
\begin{equation}
    P(\mathbf{X|C}) \propto P(\mathbf{X})  \prod_k P(C_k|\mathbf{X}),
\end{equation}
where, $P(\mathbf{X})$ is the prior distribution, and $C_k$ include the various constraints from nuclear physics and astrophysics. From the nuclear physics side, we first would like to reproduce the properties of nuclear matter around saturation, i.e. the values of the NMPs. Our model parameters should be sampled in a way such as to cover the full ranges of possible values for the NMPs. These values are given in Table~\ref{tab:prior}, which were chosen such as to cover largely their present uncertainties. In practice, we do not apply any explicit likelihood model for the NMPs, but rather consider uniform priors for the model parameters of the GDFM Lagrangian within the ranges reported in Table \ref{tab:parameters},  and then select models within the uncertainty of the NMPs as assumed in Table~\ref{tab:prior}.

\begin{table}
    \centering
    \begin{tabular}{c|c|c}
        \hline 
        \hline
      Parameters & Maximum value & Minimum value \\
         \hline 
        \hline
        $a_\sigma$ & 10.295748 & 6.9837231 \\
        $b_\sigma$ & 3.2618188 & 2.0238622 \\
        $c_\sigma$ & 2.7911622 & 1.6943625 \\
        $d_\sigma$ & 5.2779045 & 2.4805772 \\
        $a_\omega$ & 13.6596588 & 9.1064392 \\
        $b_\omega$ & 2.35939872 & 1.57293248 \\
        $c_\omega$ & 8.2559356 & 5.0097963 \\
        $d_\omega$ & 1.6719065 &  0.67148104 \\
        $a_\rho$ & 1.0 & -1.0 \\
        $b_\rho$ & 7.312709592 & 4.875139728 \\
        $c_\rho$ & 0.66405387 & 0.40285884 \\
        $d_\rho$ & 1.2092027 & -1.2112768 \\
        \hline 
        \hline
    \end{tabular}
    \caption{ Ranges of model parameters used to explore the NMPs}
     \label{tab:parameters}
\end{table}


The filters that an EOS must satisfy can be divided broadly in two groups, the low-density (LD) filters, {originating from nuclear theory and experiments,} and the high-density (HD) {ones, mainly inferred by astronomical observations}. We have used the theoretical calculations from $\chi$-EFT for SNM and  pure neutron matter (PNM) \cite{Drischler:2015eba} in the range of baryon number densities from 0.02 to 0.16 fm$^{-3}$ as our {main} LD filter. This theoretical band is interpreted as a
90\% confidence interval, and for this reason it is enlarged by 10\% on each side. We use this constraint as a band-pass type filter, meaning that we only considered EOS models that pass through the uncertainty region predicted by the calculations. Since this is a range predicted from different theoretical model Hamiltonians for the $\chi$-EFT, we have not interpreted this uncertainty region as distributed as Gaussian. {An extra constraint, particularly acting  on the low-order isoscalar NMPs, comes from nuclear experimental data. Indeed,} upon constructing the crust EOS, the parameters entering the CLDM model in addition to the NMPs are optimized with the AME2016 mass table. 
The likelihood function with the LD filter then takes the form,
\begin{equation} \label{eq:LD}
    P_{LD}(\mathbf{C}_{\text{nuclear}}|\mathbf{X}) \propto \omega_{\chi \text{-EFT}}(\mathbf{X})  P_{\text{AME}}(\mathbf{X}),
\end{equation}
where $\omega_{\chi \text{-EFT}}$ is either $0$ or $1$, depending whether the EOS model passes through the band predicted by $\chi$-EFT or not. 

Our HD filter comes from the astrophysical observations of massive pulsars, and combined tidal deformability ($\tilde{\Lambda} $) from GW170817. In particular, we use the mass-measurement of J0348+0432 as $2.01 \pm 0.04 M_\odot$ reported by Antoniadis et. al \cite{Antoniadis:2013pzd} as a Gaussian likelihood, 
\begin{equation}
    P(\mathbf{C}_{M_{\text{max}}}|\mathbf{X}) = \frac{1}{2} \left[ 1 + erf\left(\frac{M_{\text{max}}(\mathbf{X})/M_\odot - 2.04}{0.04\sqrt{2}} \right) \right].
    \label{mmax}
\end{equation}
Then, we impose the constraints on $\tilde{\Lambda}$ from Ref.~\cite{LIGOScientific:2018hze}. The value of $\tilde{\Lambda}$ thereby depends on several other parameters of the binary system notably the mass ratio $q$ and the chirp mass. The latter has been determined rather precisely to be ${\cal M}_{\text{chirp}} = 1.186 \pm 0.001 M_\odot$. We construct the likelihood for the data from GW170817 assuming the low-spin prior as,
\begin{equation}
    P(\mathbf{C}_{\text{LVC}}|\mathbf{X}) = \sum_i P_{\text{LVC}}(\tilde{\Lambda}[q^i],q^i).
    \label{lvc}
\end{equation}
For each EOS, we vary $q$ within [0.73,1.00], to generate a binary with ${\cal M}_{\text{chirp}} = 1.186$, corresponding $\tilde{\Lambda}$ and then find the probability from Eq. \eqref{lvc}. The likelihood function with the HD filter is simply the product,

\begin{equation}
    P_{HD}(\mathbf{C}_{\text{astro}}|\mathbf{X}) = P(\mathbf{C}_{M_{\text{max}}}|\mathbf{X}) \times P(\mathbf{C}_{\text{LVC}}|\mathbf{X})
\end{equation}
{An extra constraint comes from the simultaneous mass-radius measurements of PSR J0030+0451 and PSR J0740+6620 by NICER.} \cite{Miller:2019cac,Miller:2021qha, Riley:2019yda, Riley:2021pdl}. {However, we have checked that } all models having passed the different filters of our choice are already compatible with NICER data {with comparable likelihood, see also \cite{Thi:2021jhz}. Therefore, we have not used this extra constraint in the HD filter}. However, our predictions will be compared to the NICER data in Section \ref{sec:properties}.
\begin{figure}
    \centering
    \begin{tabular}{cc}
        \includegraphics[width=0.50\textwidth]{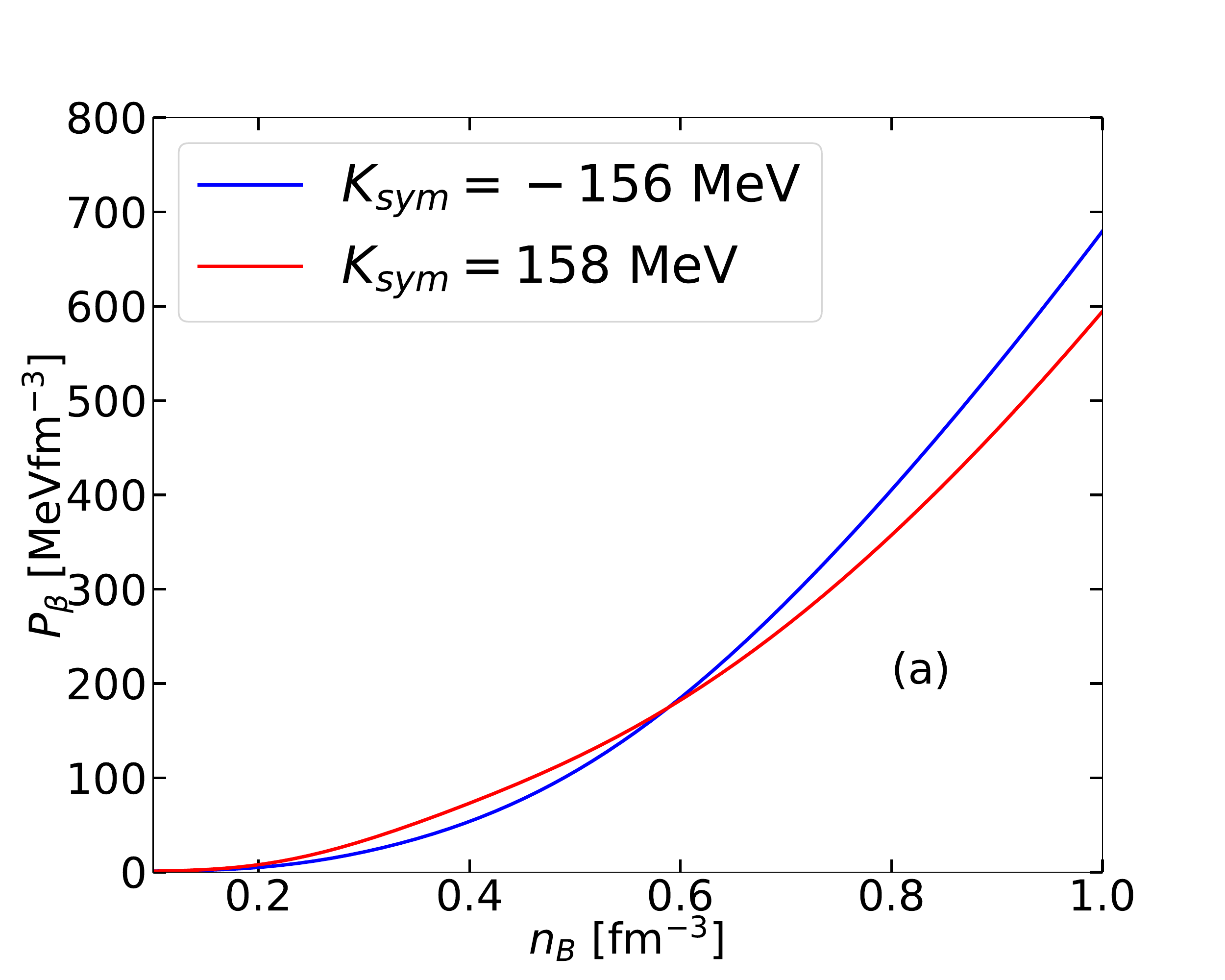} &  \includegraphics[width=0.50\textwidth]{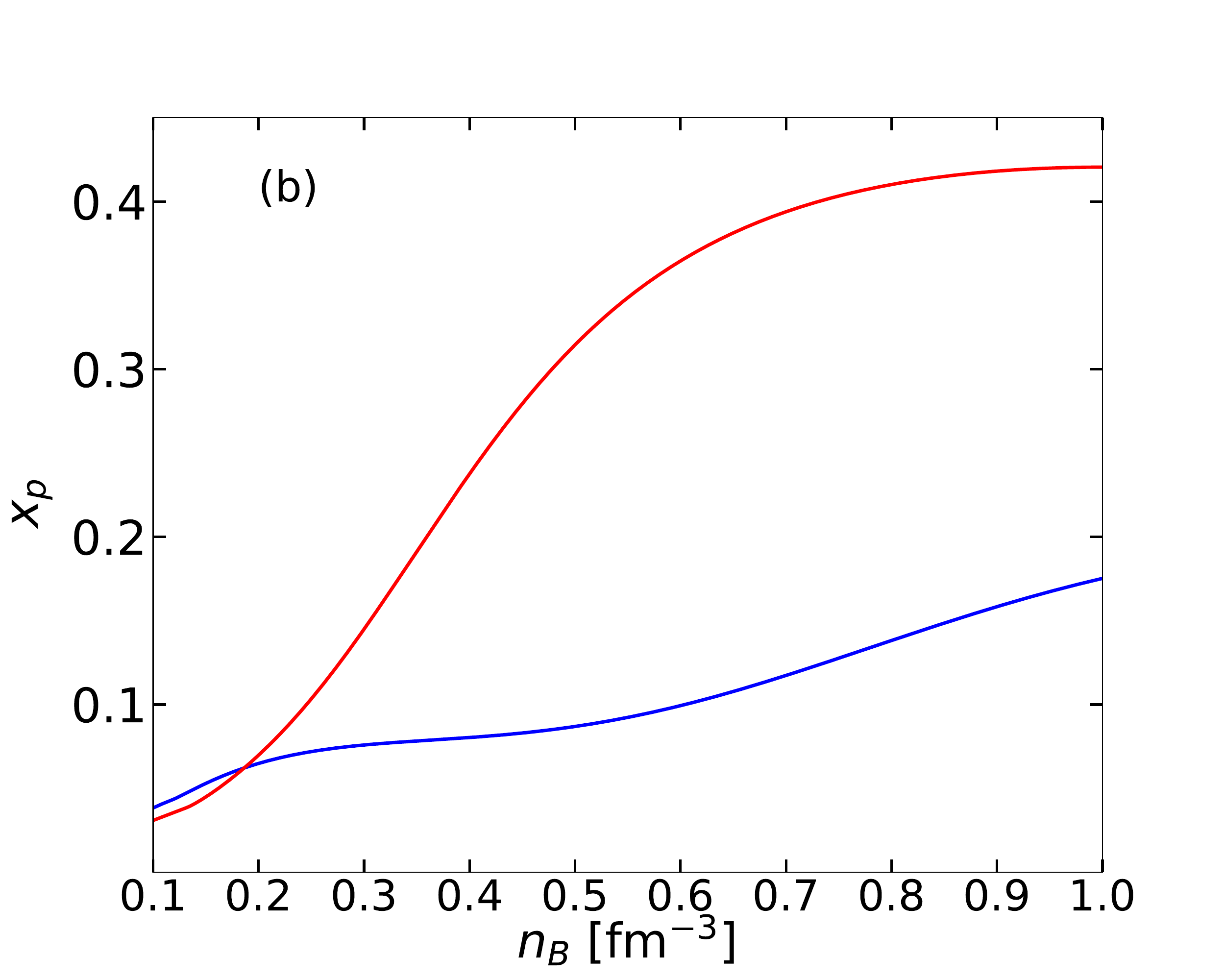} \\
        \includegraphics[width=0.48\textwidth]{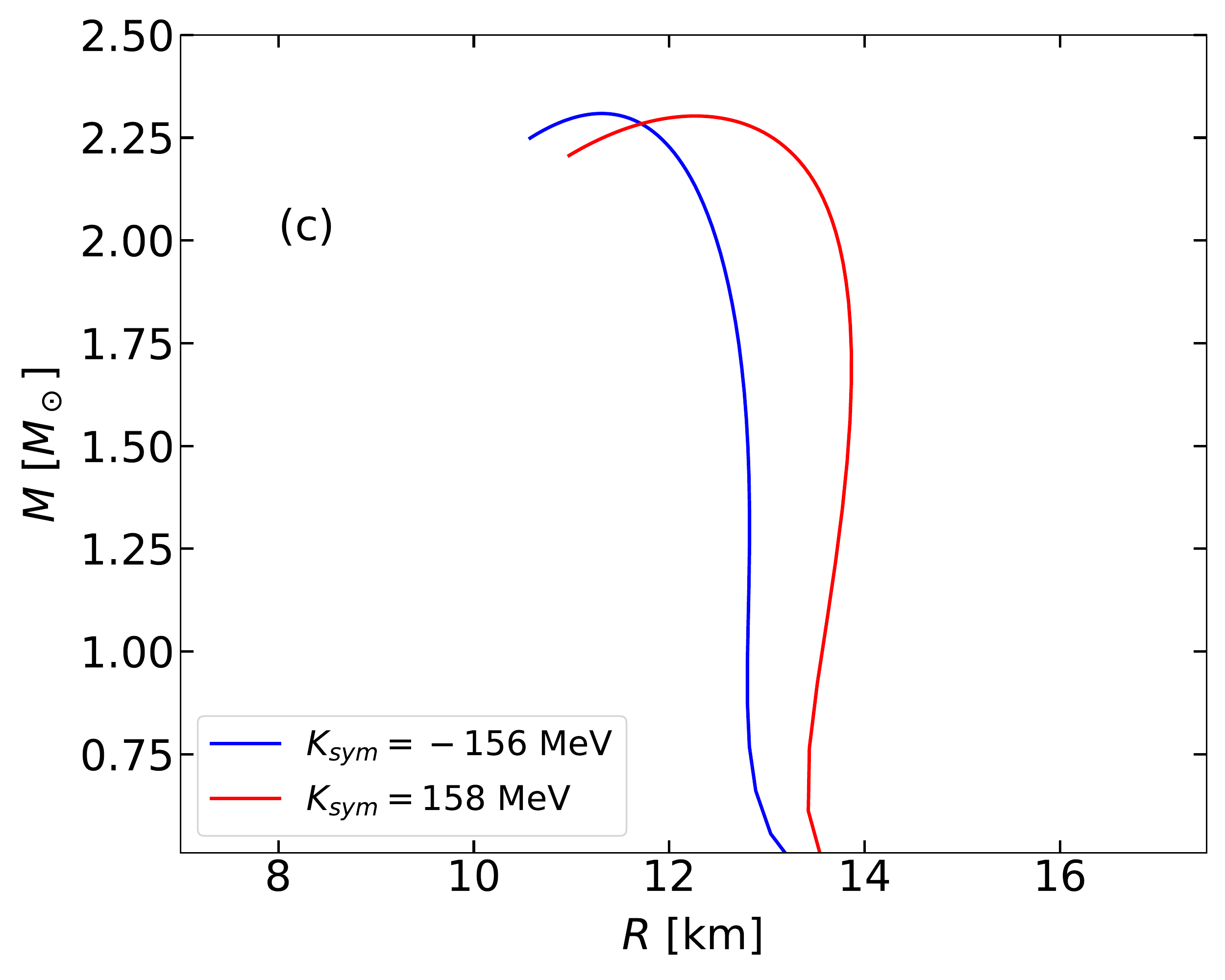} & \includegraphics[width=0.48\textwidth]{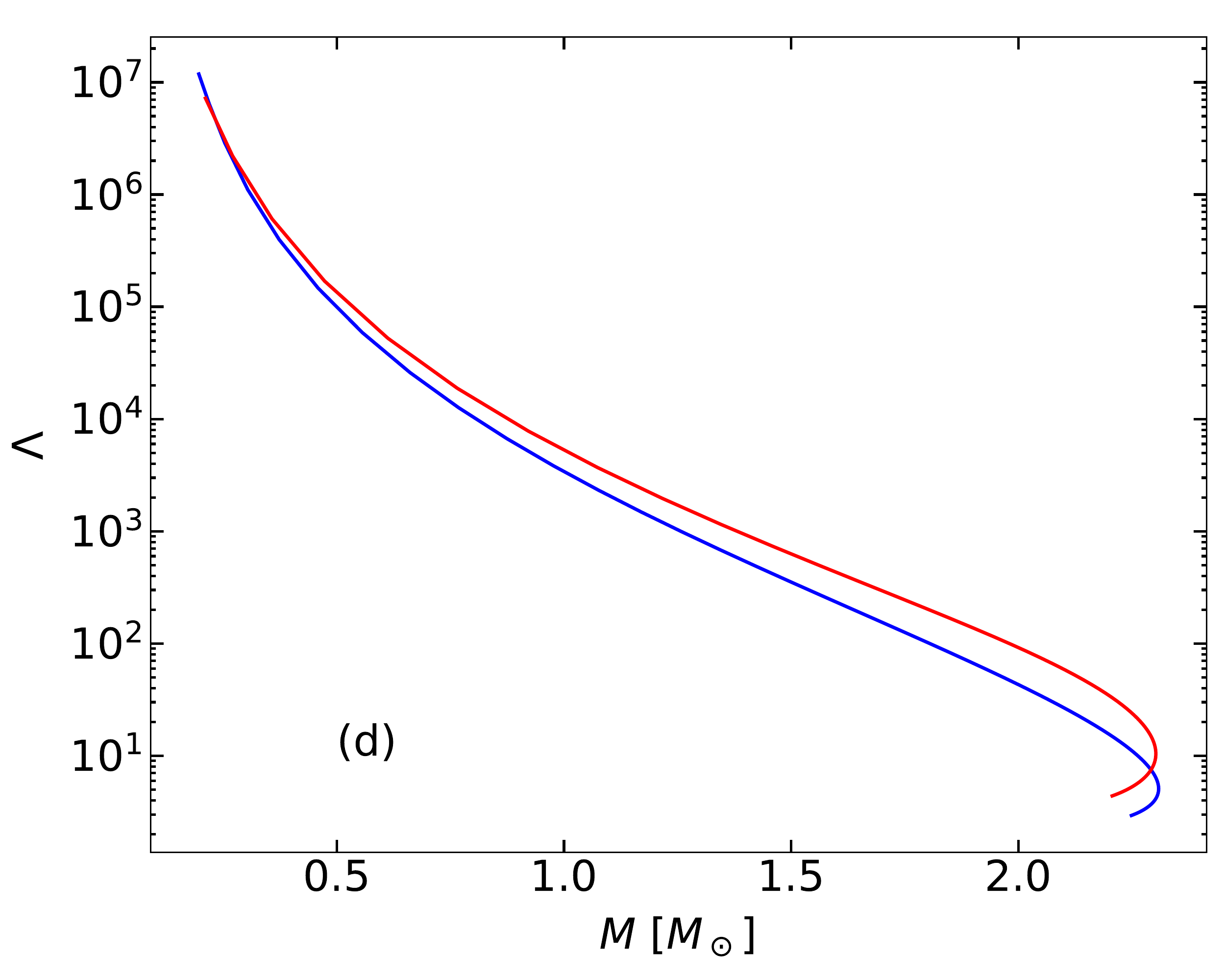}
    \end{tabular}
   \caption{Pressure (panel a) and  proton fraction (panel b) as a function of baryon density $n_B$, mass-radius (panel c), and tidal deformability (panel d) for two example EOS models with negative and positive $K_{sym}$, respectively, see text for details.
   }
    \label{compare_ksym_eos}
\end{figure}
\section{Results}
\label{results}
\subsection{Nuclear matter parameters}
\label{ssec:NMPs}
Let us start the discussion of our results with the NMPs of the GDFM model in terms of the model parameters $a_j, b_j, c_j, d_j, j=1,\dots,4$. As mentioned before, our choice of the functional form for the density dependence of the couplings allows to well explore the isovector interaction channels. In particular, we manage to explore regions of positive $K_{sym}$ that have not been studied earlier within this type of models, see e.g. the recent studies~\cite{Malik:2018zcf,Beznogov:2022rri,Malik:2022zol,Malik:2023mnx}. To better motivate this aspect of our work, let us present two example EOS models: one with $K_{sym} = 158$ MeV and the other one with $K_{sym} = -156$ MeV. The lower order NMPs for these two sets are very similar, i.e. they have similar symmetry energy and incompressibilities. The model parameters corresponding to these two sets are provided in Table~\ref{compare_ksym}. We construct the EOS models in $\beta$-equilibrium and then calculate the TOV sequences, shown in Fig.~\ref{compare_ksym_eos}. One may see that the NS EOS as well as {the maximum mass obtained from} their TOV solutions are very similar. The tidal deformabilities as function of NS mass can differ by up to factor 2-3 for high NS masses, but this difference is not significant in view of the current precision on $\Lambda$ from GW detections.  {For a given mass, the radii are slightly higher for the $K_{sym}>0$ case, as expected for a stiffer equation of state }.  However, the proton fraction looks completely different between the two models. Positive $K_{sym}$ results in a much larger proton fraction thus drastically changing the chemical composition of the star. {This latter determines the transport properties of the star, such as the thermal and electrical conductivity or neutrino emissivity. This information is difficult to extract directly from the astronomical observations sensitive to the EOS; still, the knowledge of the composition is crucial to correctly model astrophysical phenomena such as NS cooling or pulsar glitches \cite{Oertel:2016bki,Potekhin:2017ufy, Montoli:2018fqz, Beloin:2018fyp, Lunney:2020gjj}.} Therefore, one can conclude that our approach provides enough flexibility to explore all possibilities for the neutron star interior structure given a typical observation of NS global properties. 


\begin{table}

   \centering
    \begin{tabular}{c|c|c}
        \hline 
        \hline
        Parameters & Model I 
        & Model II 
        \\
        \hline 
        \hline
        $a_\sigma$ & 8.225494  & 7.849499\\
        $b_\sigma$ & 2.7079569 & 2.9940229 \\
        $c_\sigma$ & 2.4776689 & 2.4904544\\
        $d_\sigma$ &  3.8630221 & 4.5836619\\
        $a_\omega$ & 10.426752 & 9.9826068\\
        $b_\omega$ & 1.6468675 & 1.7422104\\
        $c_\omega$ &  6.8349408 & 7.4337329\\
        $d_\omega$ &  1.4458185 & 1.3409159\\
        $a_\rho$ & 0.64584657 & -0.85630723 \\
        $b_\rho$ & 5.2033131 & 6.7966817 \\
        $c_\rho$ & 0.4262597 & 0.517707 \\
        $d_\rho$ & -0.1824181 & 1.0008296 \\ 
        \hline
        $ n_{sat}$ (fm$^{-3}$)  & 0.16194209 & 0.16249048 \\
        $m^*$ & 0.67879492 & 0.69872797 \\
         $ E_{sat}$ (MeV) & -15.526417 & -15.082767\\
         $ K_{sat}$ (MeV) & 249.10229 & 236.40272 \\
         $ E_{sym}$ (MeV) & 32.908066 & 31.465153 \\
         $ L_{sym}$ (MeV) & 53.129645 & 76.468950\\
        $K_{sym}$ (MeV) & -156.06294 &  158.18981\\
        \hline 
        $M_{max}$ ($M_\odot$) & 2.30883 & 2.30272 \\
        $R_{M_{max}}$ (km) & 11.3041 & 12.2781 \\
        $R_{1.4}$ (km) & 12.8175 & 13.7942 \\
        \hline  
        \hline 
    \end{tabular}
    \caption{Two representative models corresponding to negative (model I) and positive (model II) $K_{sym}$}
    \label{compare_ksym}
\end{table}

\begin{figure}
    \centering
    \includegraphics[width=1.0\textwidth,height=0.6\textheight]{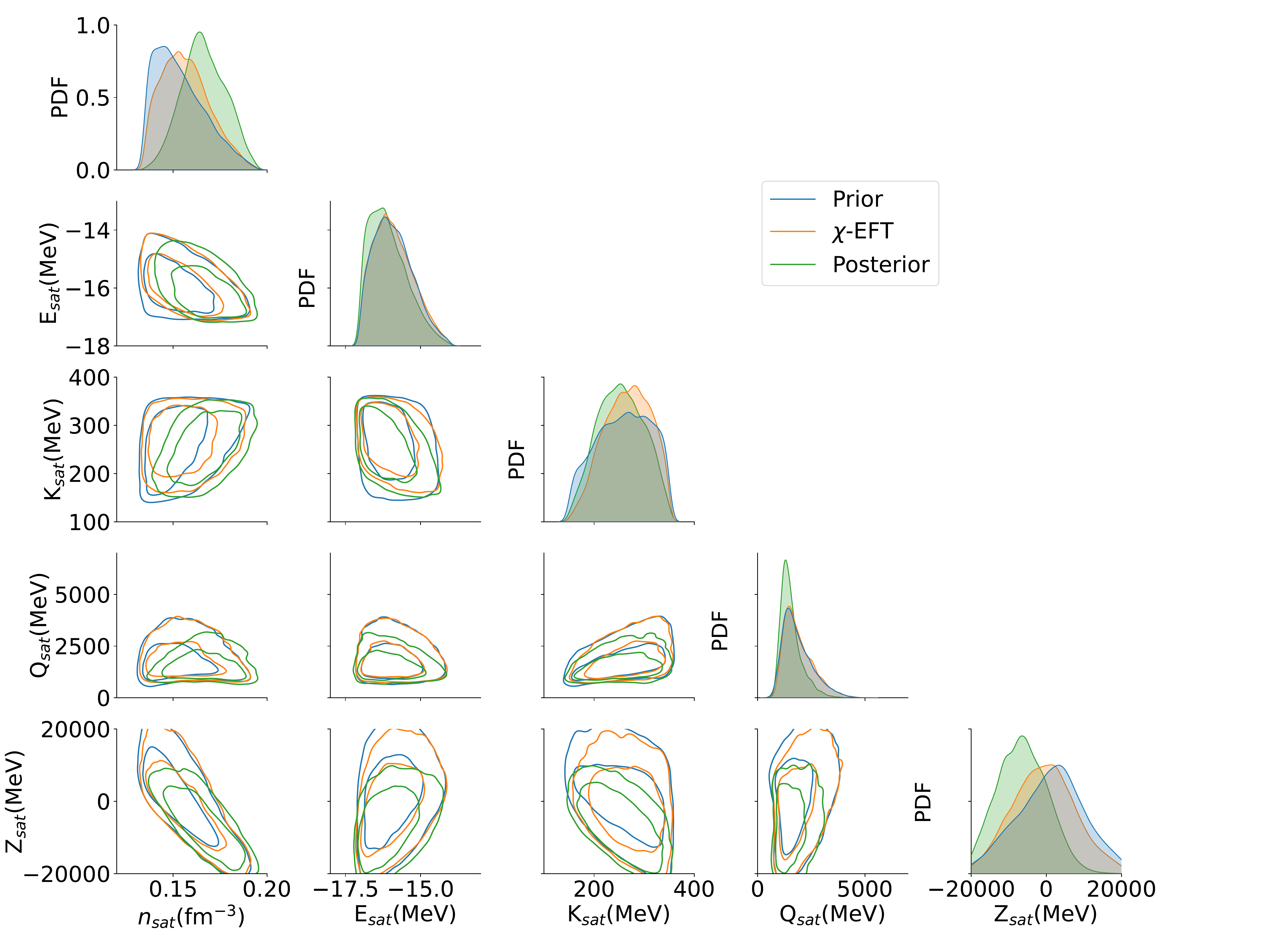}
    \caption{Isoscalar NMPs for the relativistic meta-model, with priors taken from the uniform paraneter distribution of the GDFM model, see Table~\ref{tab:parameters}. {The prior (blue), low-density (denoted as $\chi$-EFT) (orange) and full (LD+HD, green) posteriors are shown. 68\% and 95\% probability contours are given in the correlation plots. The green distributions are very close to the orange ones, and almost invisible on the scale of the figure.}}
    \label{fig:isoscalar} 
\end{figure}
\begin{figure}
    \centering
    \includegraphics[width=1.0\textwidth,height=0.6\textheight]{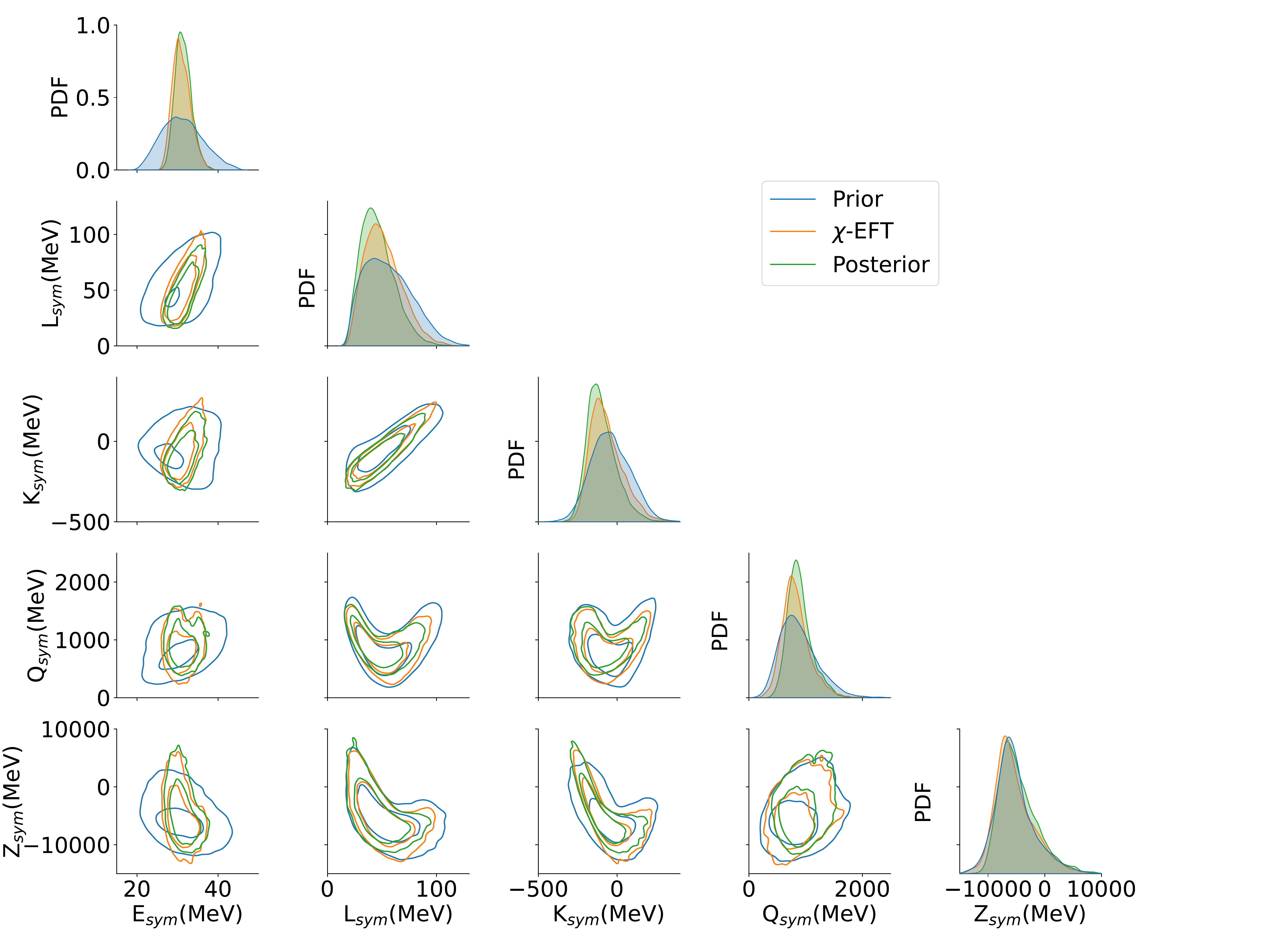}
    \caption{Same as Fig. \ref{fig:isoscalar} but for the isovector NMPs.}
    \label{fig:isovector}
\end{figure}
Next, we explore further the range of values for the NMPs produced by the relativistic meta-model with the samples uniformly drawn from the range of model parameters shown in Table~\ref{tab:parameters}. In Figs.~\ref{fig:isoscalar} and \ref{fig:isovector}, we show the distributions of the isoscalar and isovector parameters respectively. {To obtain what we will call in the following our prior (blue distributions in Figs.~\ref{fig:isoscalar} and \ref{fig:isovector}), we first } 
stitch the crust EOS calculated for the same set of NMPs following \cite{Carreau:2019zdy} to the high-density GDFM EOS model. For the 3rd and 4th order NMPs, we do not assume them to be the same for low-density and high-density, i.e. crust and core. The parameters named $Q$'s and $Z$'s correspond to those calculated from the GDFM calculations, while independent third and fourth-order parameters are generated for the calculations of the crust EOS \cite{Mondal:2022cva}. {The construction of the crustal EoS further requires the definition of the surface parameters that are optimized for each EoS set, see  \textcite{Carreau:2019zdy} for details. This is done imposing the reproduction of the nuclear masses 
($P_{\text{AME}}$), and further the models unable to produce a viable crust are removed from our samples. As a consequence, a part of the impact of the nuclear measurements is already introduced effectively in what we call ``Prior". }

After application of the extra low-density contraints via the $\omega_{\chi \text{-EFT}}$ filter (see Equation \eqref{eq:LD}), the final posteriors are found by imposing the likelihood models from the astrophysical observations as discussed in section~\ref{data}.
As we can see from Figs.~\ref{fig:isoscalar} and \ref{fig:isovector},  the $\chi$-EFT  constraint has a small effect on the isoscalar parameters, while  it is more effective in narrowing the distributions of the low-order isovector parameters, as expected.  
Globally, we can say that the filters are not very effective in determining the optimal values of the parameters. This is due to the fact that the physical properties are not determined by specific  well identified parameters, but only by complex combinations of the whole parameter set. For this reason, it is not possible to a priori reduce the prior range of the parameters as given in Tables \ref{tab:prior} and \ref{tab:parameters}  without creating an uncontrolled bias. Only $\sim 0.43$ \% of the models are retained by the pass-band filters, but the size of the prior, $N=6.5\times 10^6$ is chosen such as to guarantee the the posterior statistics is sufficient to have convergent estimations for the observables shown in the paper.
Interestingly,
the prior distribution contains some correlations between the NMPs that are sharpened by the $\chi$-EFT filter, notably the $E_{sym}$-$L_{sym}$ and the $K_{sym}$-$L_{sym}$ correlations on Fig.~\ref{fig:isovector}. This suggests that those built-in correlations of the GDFM correspond to physical properties, and are not induced by the arbitrary form of the density dependence of the meson couplings. Other more unexpected correlations are seen in the isoscalar sector, notably between $n_{sat}$ and $Z_{sat}$. Unfortunately, the existing constraints are too loose to influence $Z_{sat}$ , and therefore it is difficult to unambiguously assign a physical content to this correlation.  We can also observe that the full symmetry energy posteriors (green distributions labelled "Posterior" in  Fig. \ref{fig:isovector}), are  very similar to the distributions obtained after the application of the low- density filters only (orange distributions labelled "$\chi$-EFT" in  Fig. \ref{fig:isovector}). This underlines the fact that present astrophysical constraints are not very effective in unambiguously  pinning down the density dependence of the nuclear symmetry energy, as already observed by different authors \cite{Mondal:2021vzt,2020ApJ...899....4X}.

\begin{figure}
    \centering
    \begin{tabular}{cc}
     \includegraphics[width=0.48\textwidth]
    {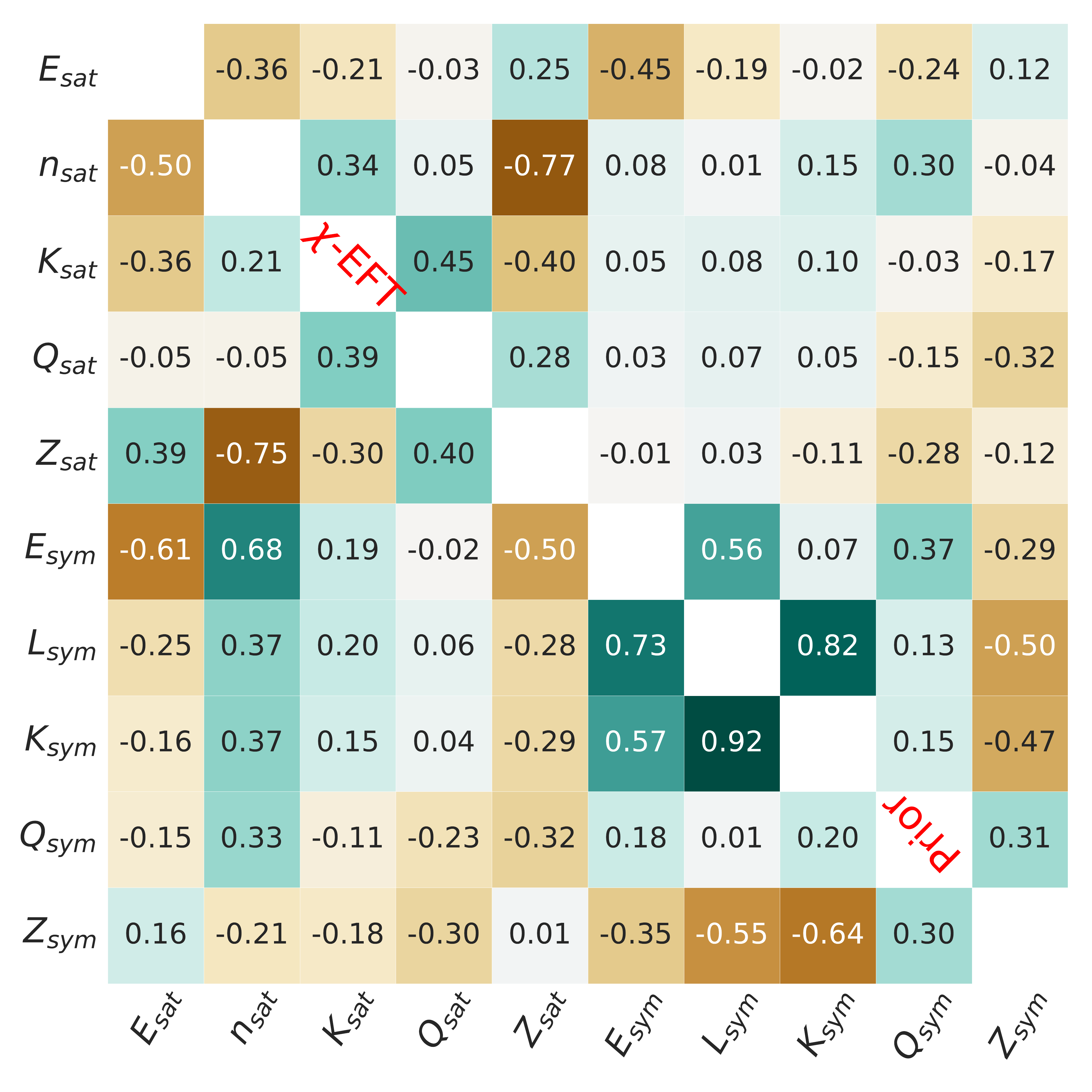}
    \includegraphics[width=0.48\textwidth]
    {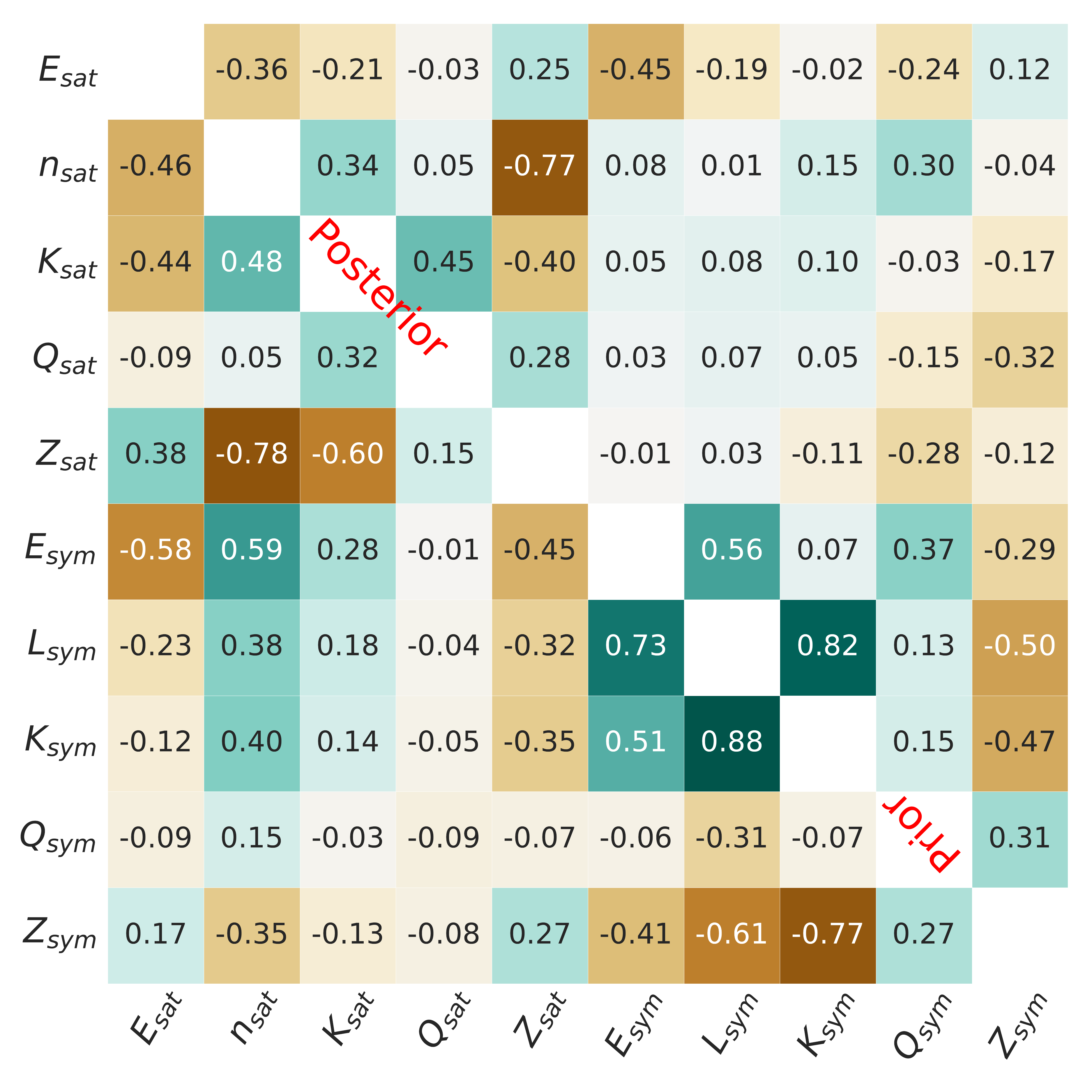}
    \end{tabular}
    \caption{{(Right): Pearson correlation coefficients between the different NMPs for the GDFM prior (upper part) and full posterior (lower part) distribution; (Left): Same as Right but for $\chi$-EFT filter in the lower part.}}
    \label{fig:corr_NMP}
\end{figure}

{ These observations can be better quantified by looking at Fig.~\ref{fig:corr_NMP}. }
In this figure, we present the Pearson correlation coefficients between the NMPs calulated from the GDFM meta-model for the priors and the posteriors. Here, we see the prior correlations as the intrinsic properties of the meta-model and find how the different filters modify them.  {We quantitatively confirm the information obtained visually from  Figs.~\ref{fig:isoscalar} and \ref{fig:isovector}).  The appearence of quasi-systematic strong correlations between the high-order parameters and the low-order ones is a very encouraging information: it suggests that more stringent constraints from nuclear data would pin down the density dependence of the EoS also at higher densities. 
}
\begin{figure}
    \centering
    \begin{tabular}{cc}
     \includegraphics[width=0.48\textwidth]
    {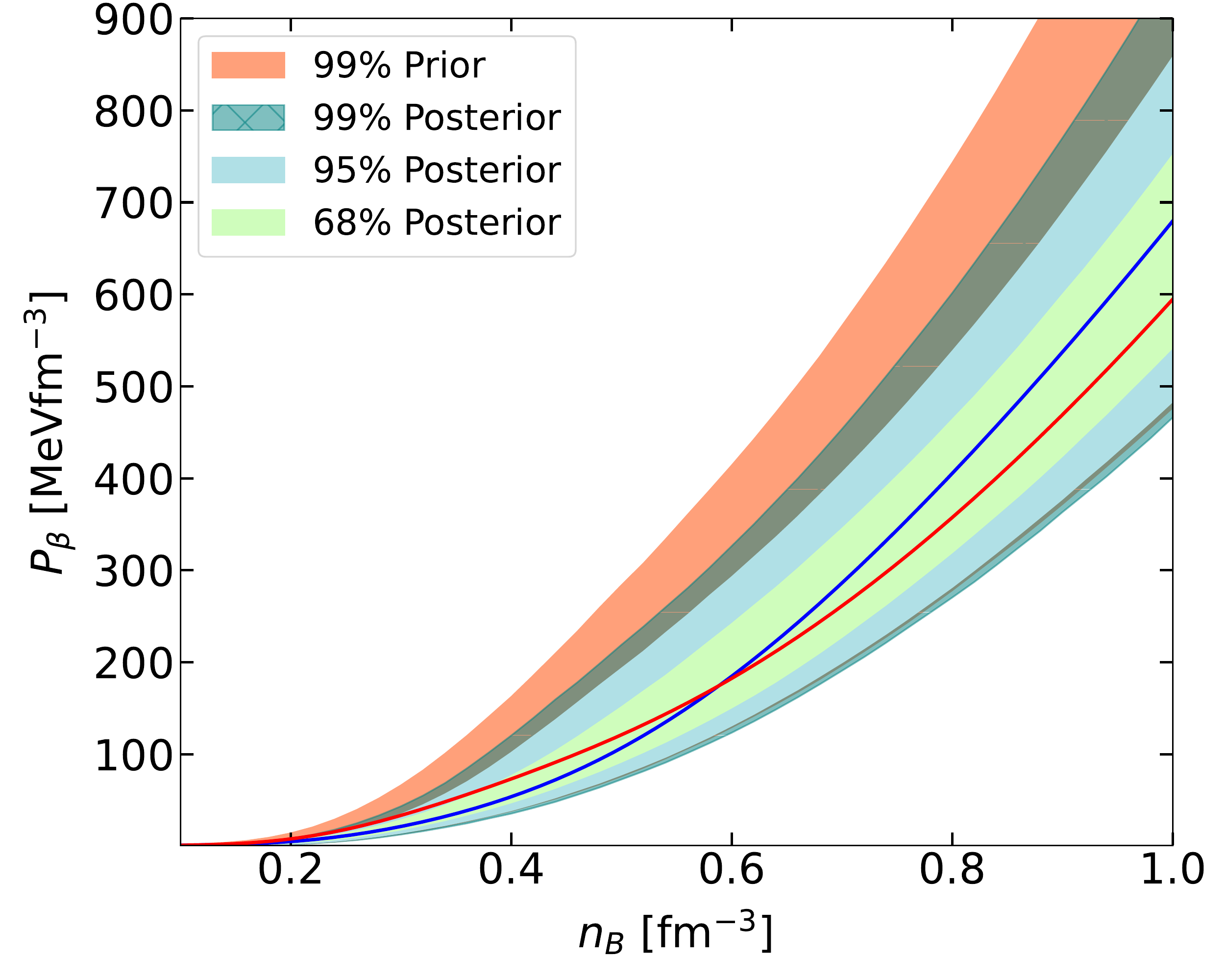}
    \includegraphics[width=0.48\textwidth]
    {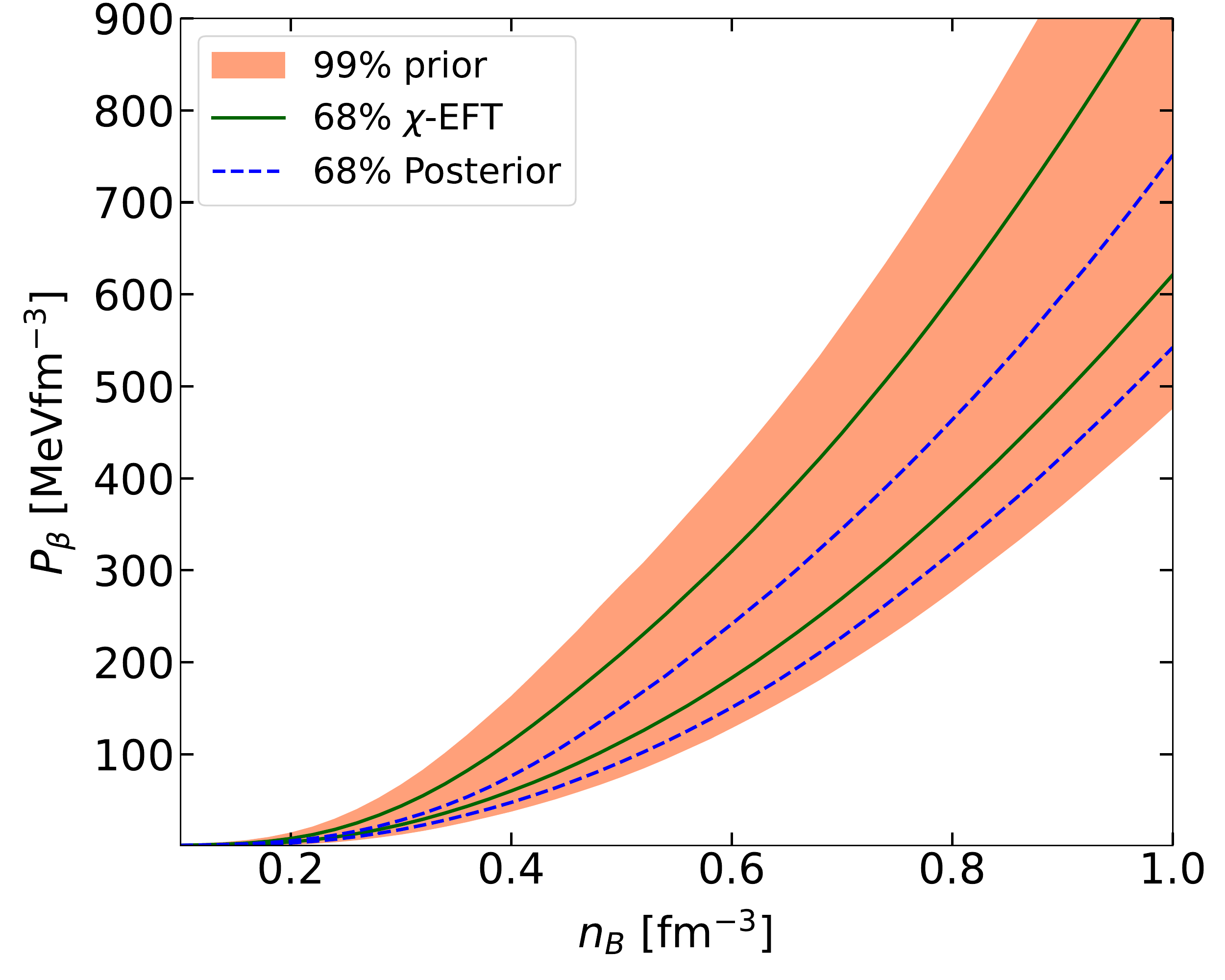}
    \end{tabular}
    \caption{Pressure at $\beta$-equilibrium as a function of baryon density. {(Left): prior and posterior distributions within different probability ranges. The blue (red) curve corresponds to the representative model I (II), respectively; (Right): the 99\% prior and the 68\% posterior are confronted to the posterior obtained applying only the low-density constraints.  The statistics is the same in each density bin.} }
    \label{fig:p-rho}
\end{figure}
\subsection{Equation of state and composition of neutron star matter}
\label{ssec:EOS}
In Fig.~\ref{fig:p-rho}, we show the pressure of cold matter in $\beta$-equilibrium as a function of the baryon number density corresponding to the NMP distributions of Figs.~\ref{fig:isoscalar} and \ref{fig:isovector}. Here, we distinguish between the EOS models generated by the GDFM meta-model in terms of $99\%$ prior, and the $68\%$ and $95\%$ posteriors obtained after applying the different filters, respectively. 
{The comparison between the distribution obtained applying only the low-density filter and the full posterior (at 68\% confidence interval), in the right panel of the figure, shows that the astrophysical observations, particularly the tidal polarizability measurement,  are very effective in constraining the equation of state. To be specific,}
we find that the posterior after applying the relevant high-density filters prefers relatively softer regions of the prior. The equations of state corresponding to the two representative models I and II associated to extreme values of the isovector compressibility $K_{sym}$ (see Fig.\ref{compare_ksym_eos} and table \ref{compare_ksym} ) are also reported in Fig.~\ref{fig:p-rho} (left panel). We can see that both models lie well within the higher probability region of our posterior, meaning that very different $K_{sym}$ values cannot be easily discriminated by measurements of the EOS,
at least within the present uncertainties of the astrophysical measurements. This is consistent with the similar mass-radius relation observed in  Fig.\ref{compare_ksym_eos}.  
It is important to stress that these models are covariant by construction, and therefore they have causality built-in and can be calculated at any density. 

\begin{figure}
    \centering
    \begin{tabular}{cc}
     \includegraphics[width=0.48\textwidth]
    {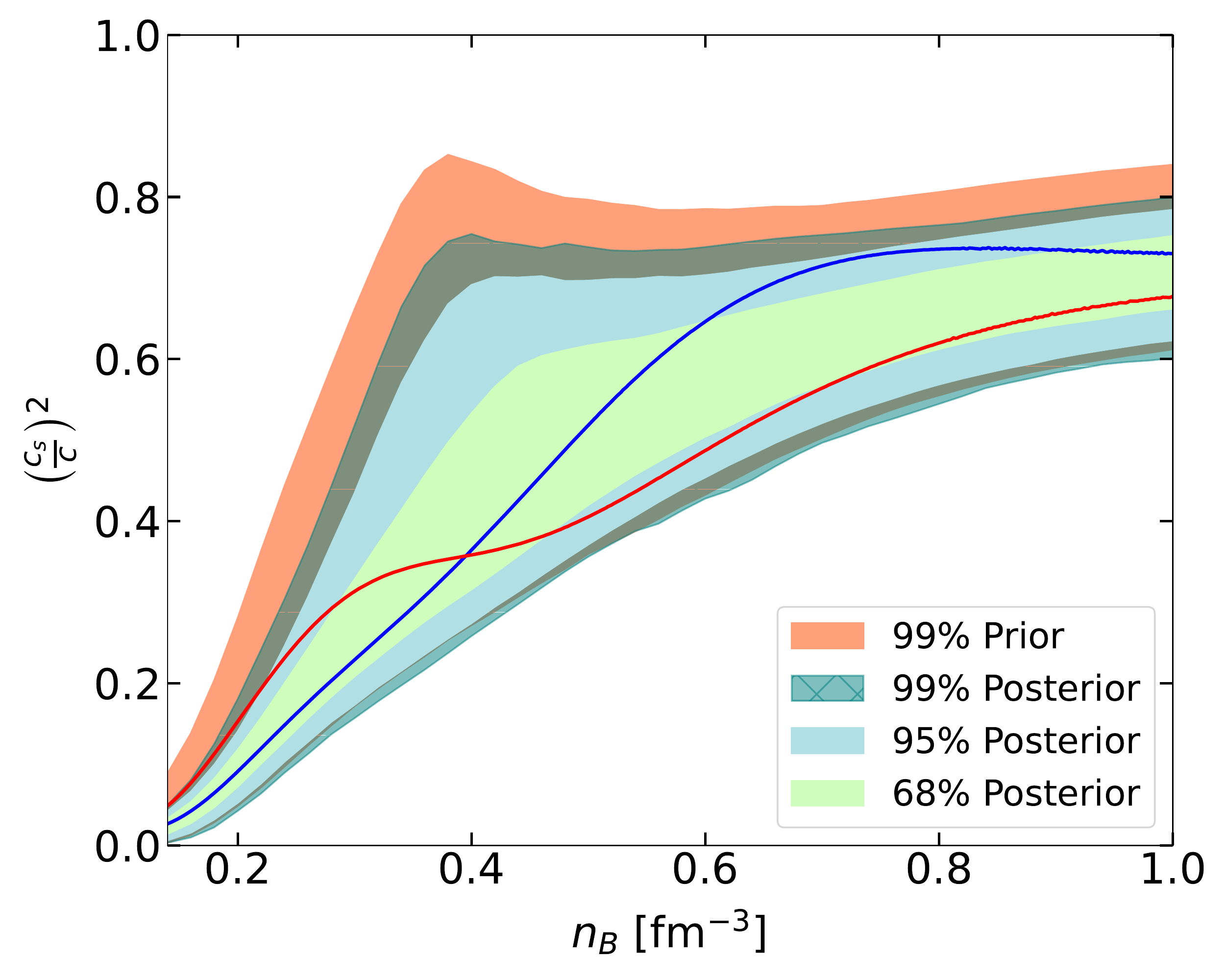}
    \includegraphics[width=0.48\textwidth]
    {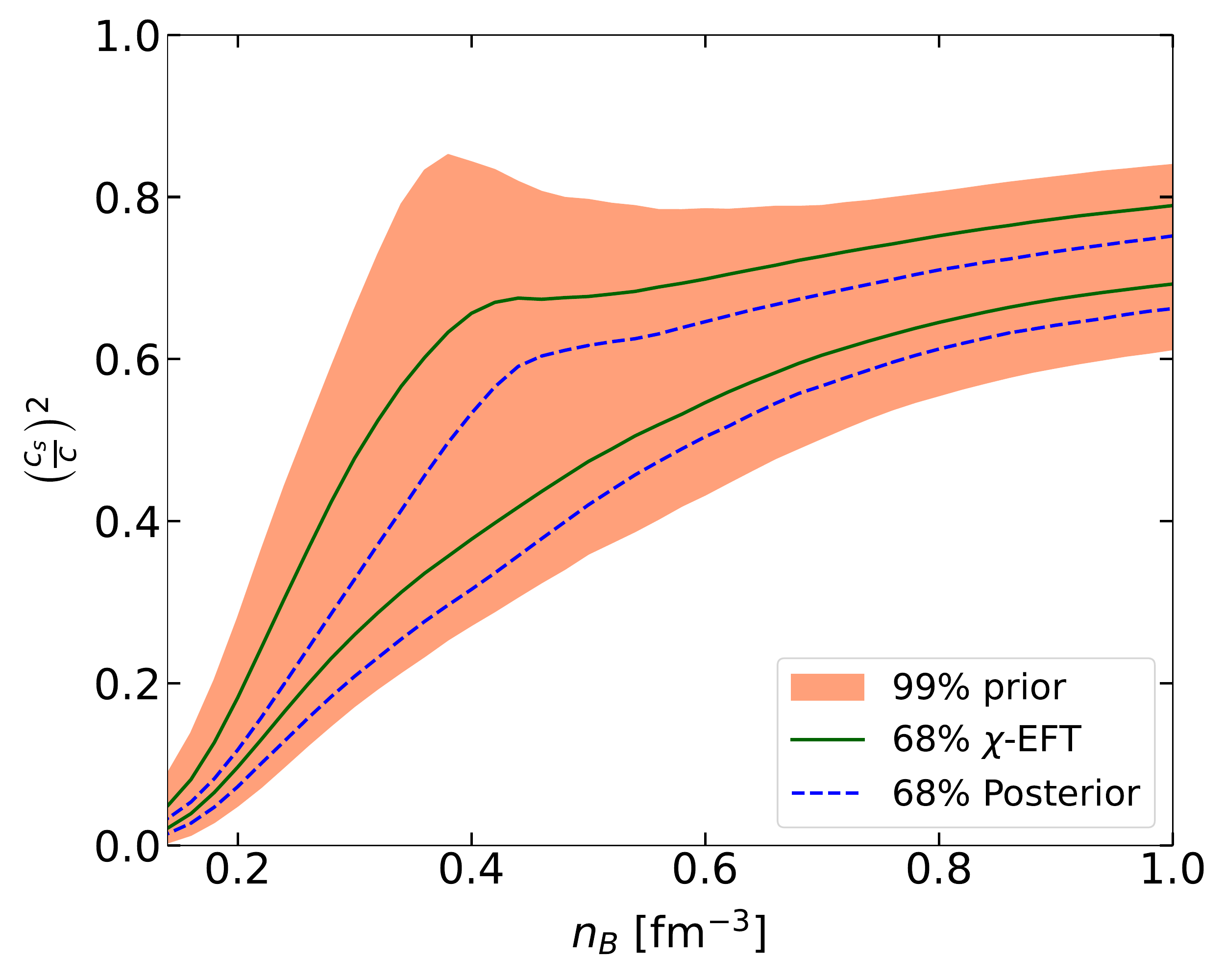}
    \end{tabular}
   \caption{Same as Figure  \ref{fig:p-rho}, but for the speed of sound as a function of the baryon number density. }
    \label{fig:cs}
\end{figure}

This is  better shown  by the behavior of the sound speed, displayed in Fig.~\ref{fig:cs} as a function of the baryon number density. As in Fig.~\ref{fig:p-rho} we show on the left panel the $99\%$ prior of the GDFM models together with the contours for the $68\%$ and $95\%$ of the posteriors.  Since this is a relativistic model, $c_s$ never reaches unity and our models remain causal. The posterior of $c_s^2$ first increases with density and then attains a roughly constant value around $\sim 0.7$ c$^{2}$. In the prior, we see a moderate kink which is not prominent in the posteriors. The most viable reason behind this behavior is that the models with stiffer EoS and thus a strong increase in the sound speed at low densities are excluded in the posteriors after applying the {likelihood of the gravitational wave measurement}, as seen in Fig.~\ref{fig:p-rho}, too. {Further, we can see from the two representative models I (blue) and II (red) associated to very different values of the $K_{sym}$ parameter, that the typical behavior of the sound speed is not necessarily monotonic, and this function can present convexities (in the $K_{sym}<0$ case), or even maxima (in the $K_{sym}>0$ case). Peaks in the sound speed are typically associated to the emergence of new degrees of freedom that soften the equation of state, see e.g. \cite{McLerran:2018hbz,Pfaff:2021kse,Shahrbaf:2021cjz,Tan:2021nat,Somasundaram:2021ljr,Malik:2023mnx,Somasundaram:2021clp} for different scenarios. Our findings demonstrate, however, that new degrees of freedom at high density are not a necessary condition for non-monotonicity in the sound speed.}{ The effect of the low-density and high-density filters are outlined in the right part of the figure.
 Consistently with the results of Fig.~\ref{fig:p-rho}, we can see that the $\chi$-EFT constraint is not very influent in determining the behavior of the sound speed, while the information from the gravitational wave observation tends to exclude the highest sound speed models.}

\begin{figure}
    \centering
    \begin{tabular}{cc}
     \includegraphics[width=0.48\textwidth]
    {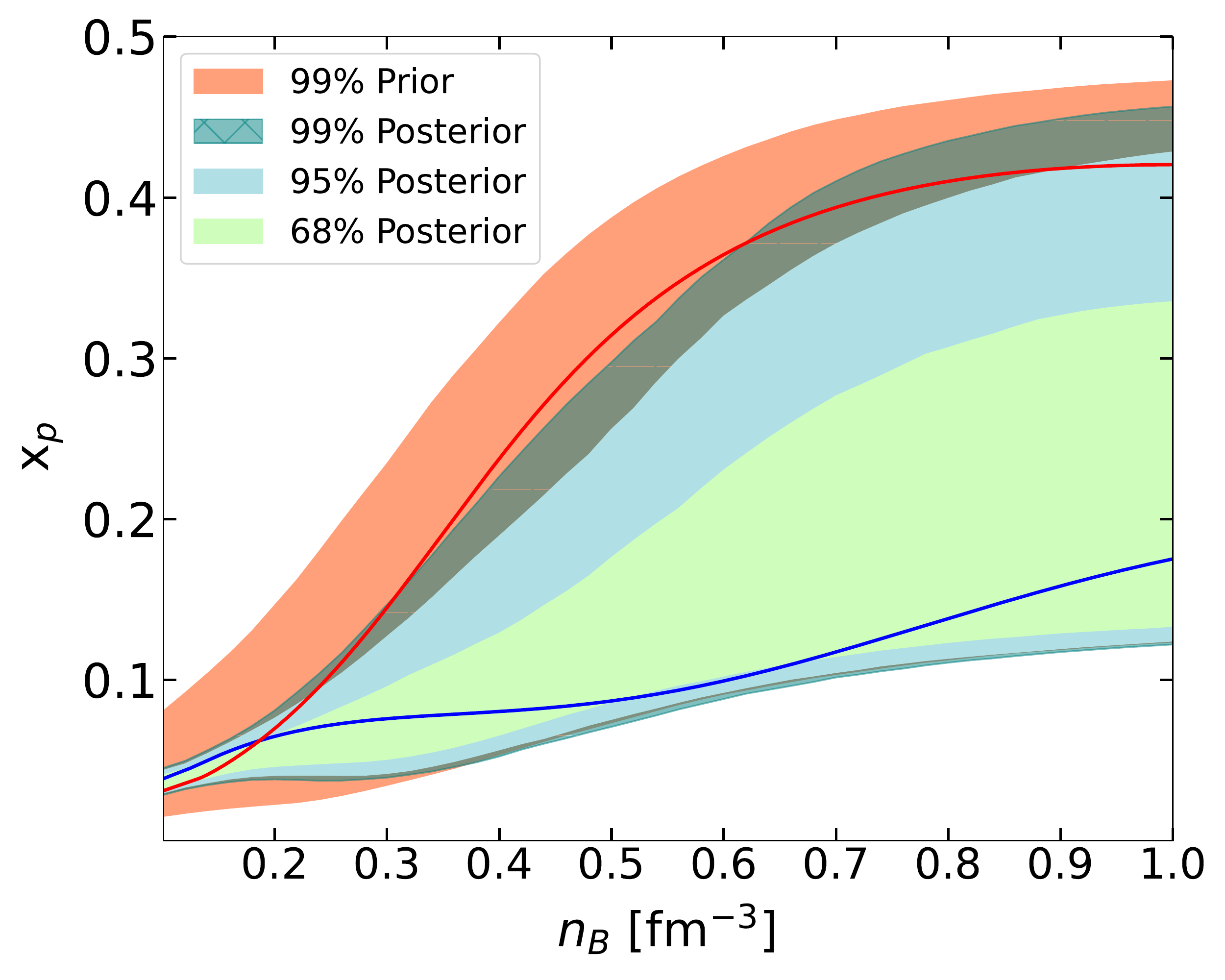}
    \includegraphics[width=0.48\textwidth]
    {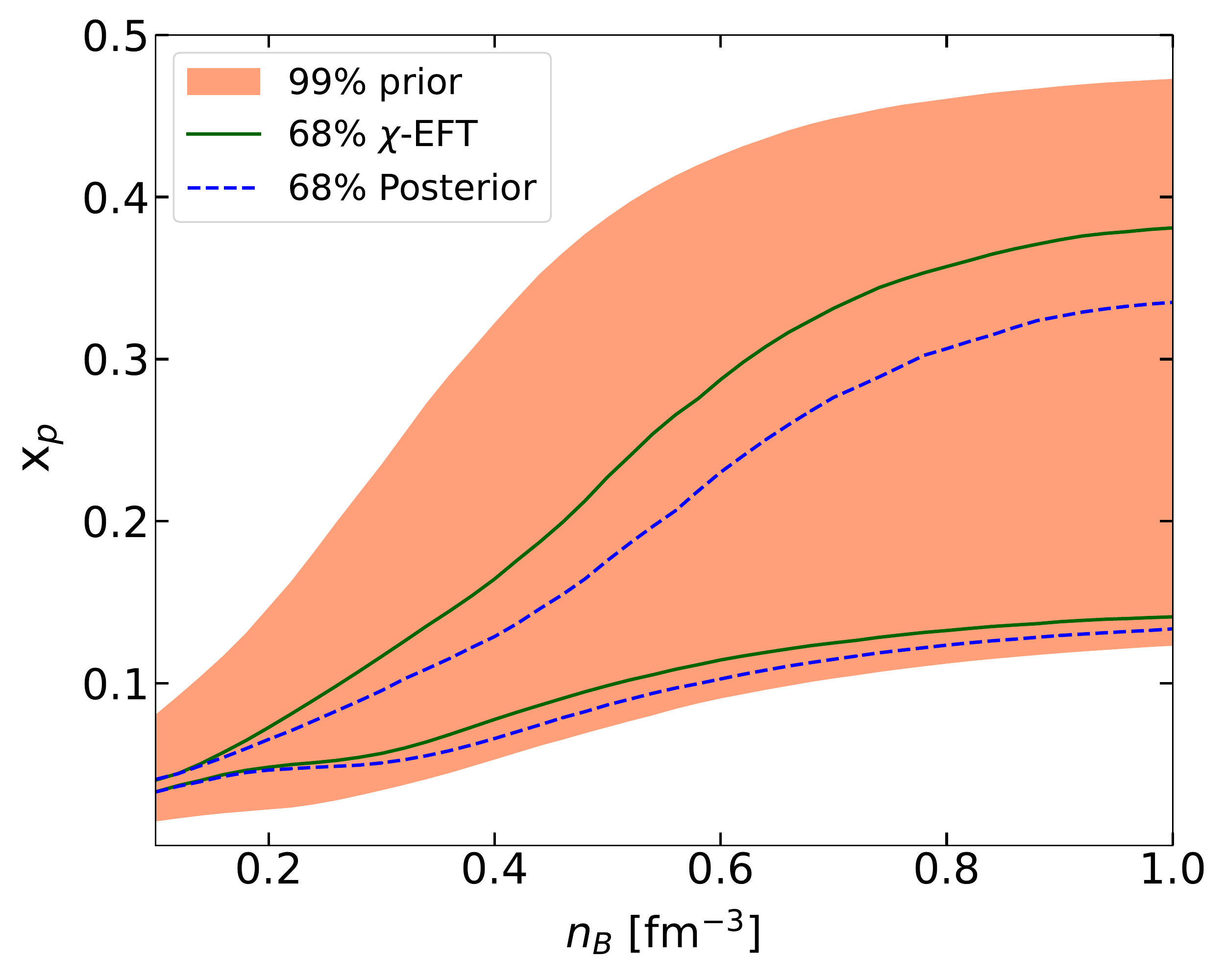}
    \end{tabular}
    \caption{Same as Figure  \ref{fig:p-rho}, but for the proton fraction as a function of baryon number density.}
    \label{fig:xp}
\end{figure}
Now, let us look at the matter composition for these different EOS models. As we have discussed before, see also Fig.~\ref{compare_ksym_eos}, the GDFM models can produce a large variation of proton fraction at a particular density due to the freedom provided in the isovector couplings, i.e. those associated with the $\rho$-meson. In Fig.~\ref{fig:xp}, we present the prior and the posterior ranges for the proton fraction of neutron star matter in $\beta$-equilibrium as function of baryon number density. 
{As in the previous figures, the separate effect of the low and high-density filters (right panel), and the representative behavior of model 1 (blue) and model 2 (red), are also reported.}
We can see that very high proton fractions reaching almost $x_p =0.5$, although being present in the prior, are excluded by the astrophysical data which push the posterior to lower proton fractions. But, due to the large freedom in the isovector interaction channel that is presently largely unconstrained, we still find that $x_p$ can reach a substantially high value at high densities, about $\sim 0.3$ considering the $68\%$ posterior and even 0.4 for the $95\%$ one. This means that   our meta-model  predicts that the core of massive NS can a priori be quite proton-rich even after all the relevant existing constraints are applied, {see also the discussion in Ref.~\cite{Providencia:2023rxc}}. In principle, detailed observations of the cooling behavior of neutron stars could help in this respect, since via the opening of direct URCA type reactions which very efficiently emit neutrinos, proton-rich neutron stars cool down very fast~\cite{Yakovlev:2000jp}. The main difficulty is, however, that observationally it is not obvious to associate a mass to a NS for which the surface temperature is measurable.  Let us stress, too, that this study supposes a purely nucleonic composition of the neutron star interior and that further uncertainties on the composition arise from the possible presence of non-nucleonic degrees of freedom at high densities, such as hyperons, mesons or a transition to deconfined quark matter.

\subsection{Neutron star properties} \label{sec:properties}
\begin{figure}
    \centering
    \begin{tabular}{cc}
     \includegraphics[width=0.48\textwidth]{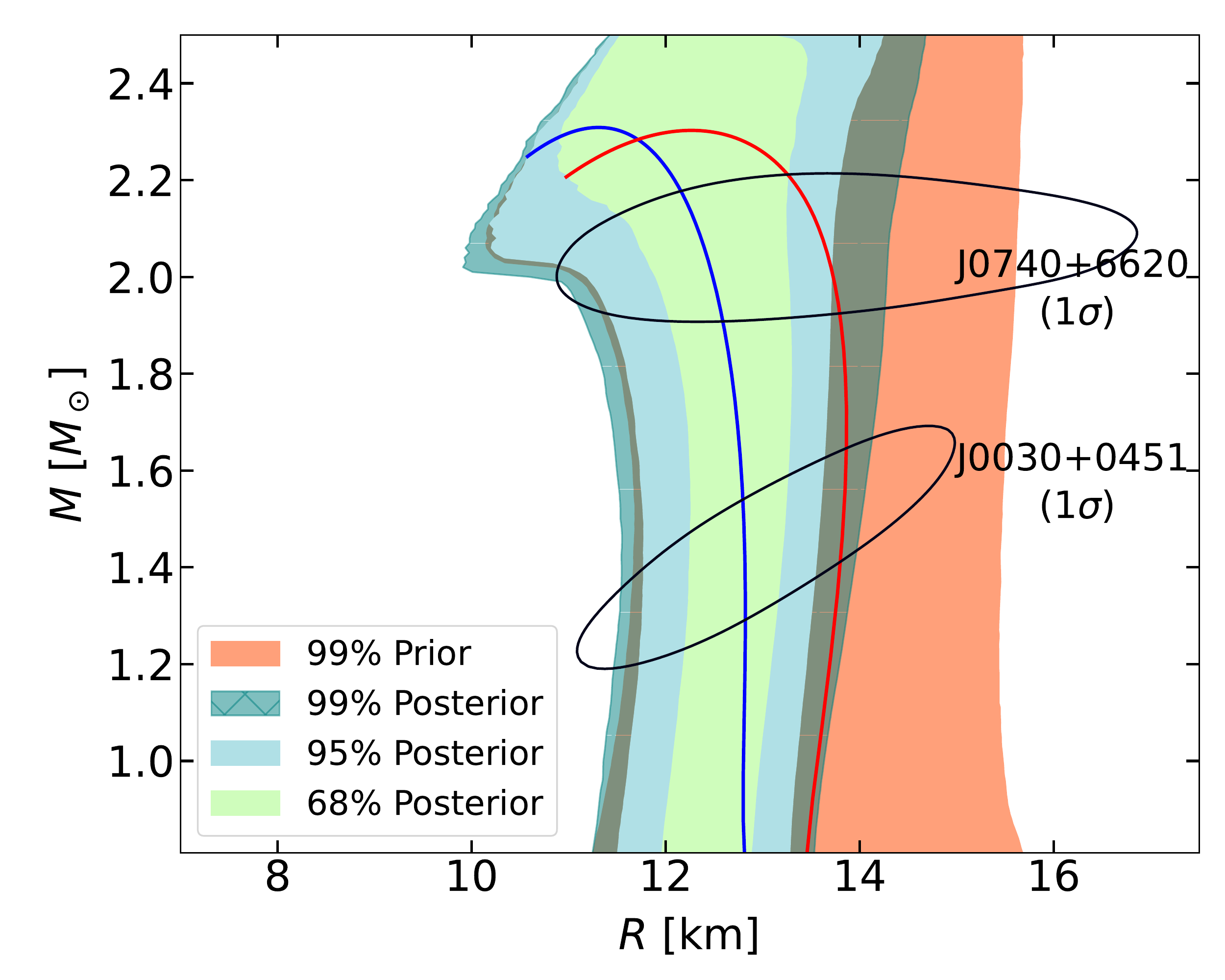}  
     \includegraphics[width=0.48\textwidth]{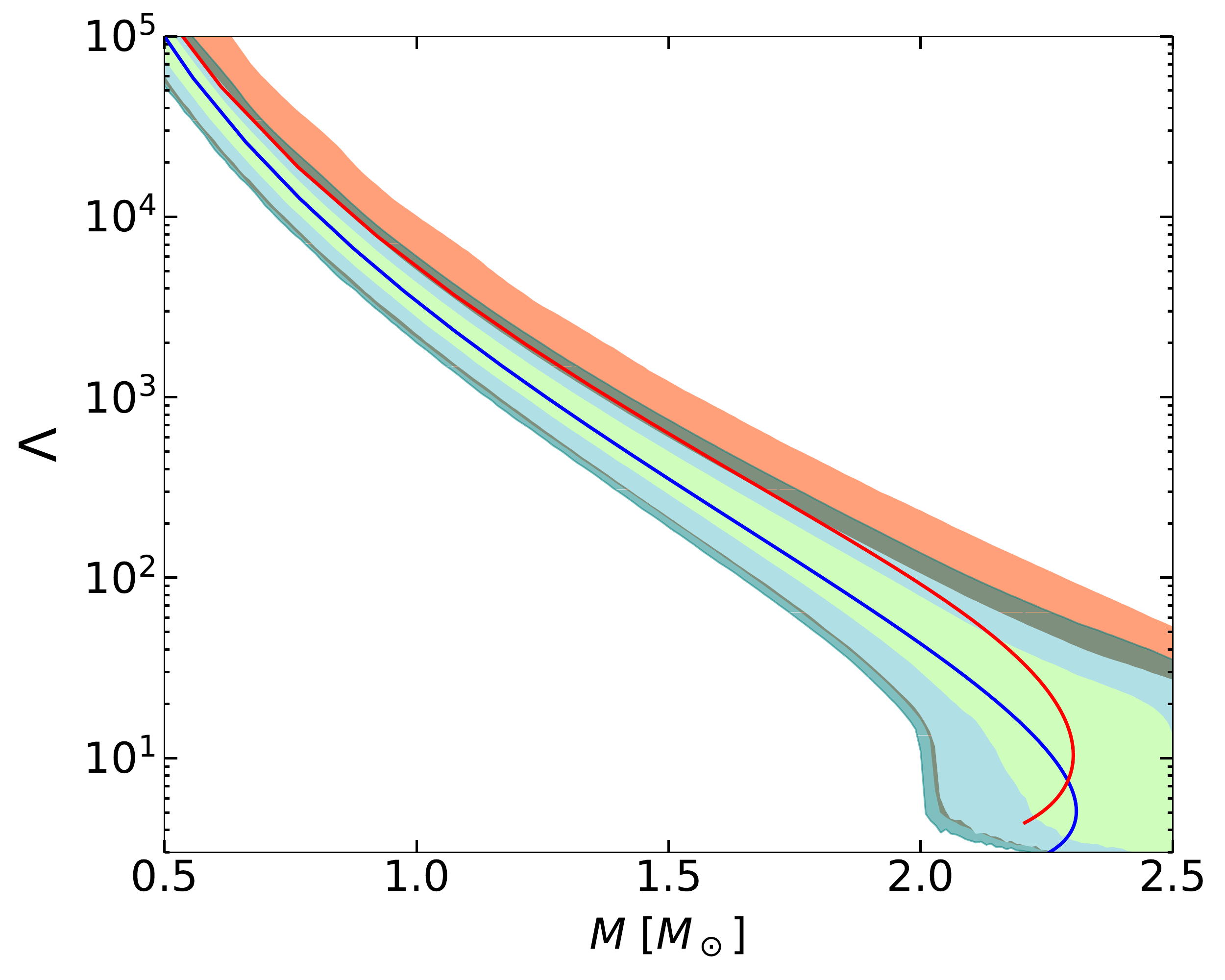}
    \end{tabular}
    \caption{Mass-radius (left) and mass-tidal deformability (right) relations corresponding to the EOS model ranges shown in Fig.~\ref{fig:p-rho} along with the Model I and II obtained in the present study. 1$\sigma$ constraints from the NICER observations are also indicated in the mass-radius panel. }
    \label{fig:mrlambda}
\end{figure}

Next, we turn our attention to the NS configurations generated from the EOS  predicted by our meta-model. To that end we solve the TOV equations~\cite{Tolman:1939jz,Oppenheimer:1939ne} for a spherically symmetric relativistic star supplemented with the equations determining the response to an external quadrupolar perturbation which allow to calculate the tidal deformabilty~\cite{Hinderer08,Hinderer09}. In Fig.~\ref{fig:mrlambda}, we display the resulting $M$-$R$ (left panel) and the $M$-$\Lambda$ (right panel) relations. Again the orange regions represent the $99\%$ prior from GDFM, whereas the posterior regions include the low-density filter from $\chi$-EFT as well as the high-density filters from massive pulsars and the tidal deformability measurement of GW170817, see Section \ref{data} for details. In the left panel, we show in addition by the two black contours the $1\sigma$ regions for the NICER measurements of J0740+6620~\cite{Riley:2021pdl,Miller:2021qha} and J0030+0451~\cite{Riley:2019yda,Miller:2019cac}, respectively. It is obvious, as mentioned before, that the GDFM is consistent with the NICER results, {which explains why adding these extra constraints would not modify our posterior distributions in an impactful way.} Following the previous trend in figures \ref{fig:p-rho} and \ref{fig:cs}, we find that the stiffer part of the prior EOS models that produce consistently larger radii are excluded. In agreement with previous results in the literature~\cite{Thi:2021jhz,Malik:2022zol}, we predict from current constraints that the radius of a $1.4 M_\odot$ star is most likely to be in the range $\sim 12-13$ km. For the tidal deformabilities, we find the equivalent results to those that are seen in the $M$-$R$ space. Median values of radius and tidal deformability for Prior, $\chi$-EFT and Posterior in 1.4$M_\odot$ and 2.0$M_\odot$ NSs, along with their 1$\sigma$ uncertainties are listed in Table \ref{tab:my_label}.

\begin{table}[]
    \centering
    \begin{tabular}{c|c|c|c|c}
    \hline
    \hline 
           & $R_{1.4}$ (km) & $R_{2.0}$ (km) & $\Lambda_{1.4}$ & $\Lambda_{2.0}$ \\
    \hline 
    \hline 
     Prior         & $13.64^{+0.77}_{-0.78}$ & $13.78^{+0.88}_{-0.87}$  & $943^{+315}_{-305}$ & $110^{+49}_{-49}$ \\
     $\chi$-EFT    & $13.38^{+0.58}_{-0.59}$ & $13.50^{+0.80}_{-0.79}$ & $845^{+250}_{-247}$ & $96^{+43}_{-42}$ \\
     Posterior     & $12.72^{+0.46}_{-0.46}$ & $12.58^{+0.68}_{-0.67}$ & $588^{+143}_{-139}$&  $54^{+24}_{-24}$ \\
    \hline 
    \hline
    \end{tabular}
    \caption{Median values of NS radii and tidal deformabilities for $1.4$ and $2.0$ $M_\odot$ stars along with their 1$\sigma$ uncertainties obtained for Prior, low density filter $\chi$-EFT and full postrior.}
    \label{tab:my_label}
\end{table}

\begin{figure}
    \centering
    \includegraphics[width=1.0\textwidth]{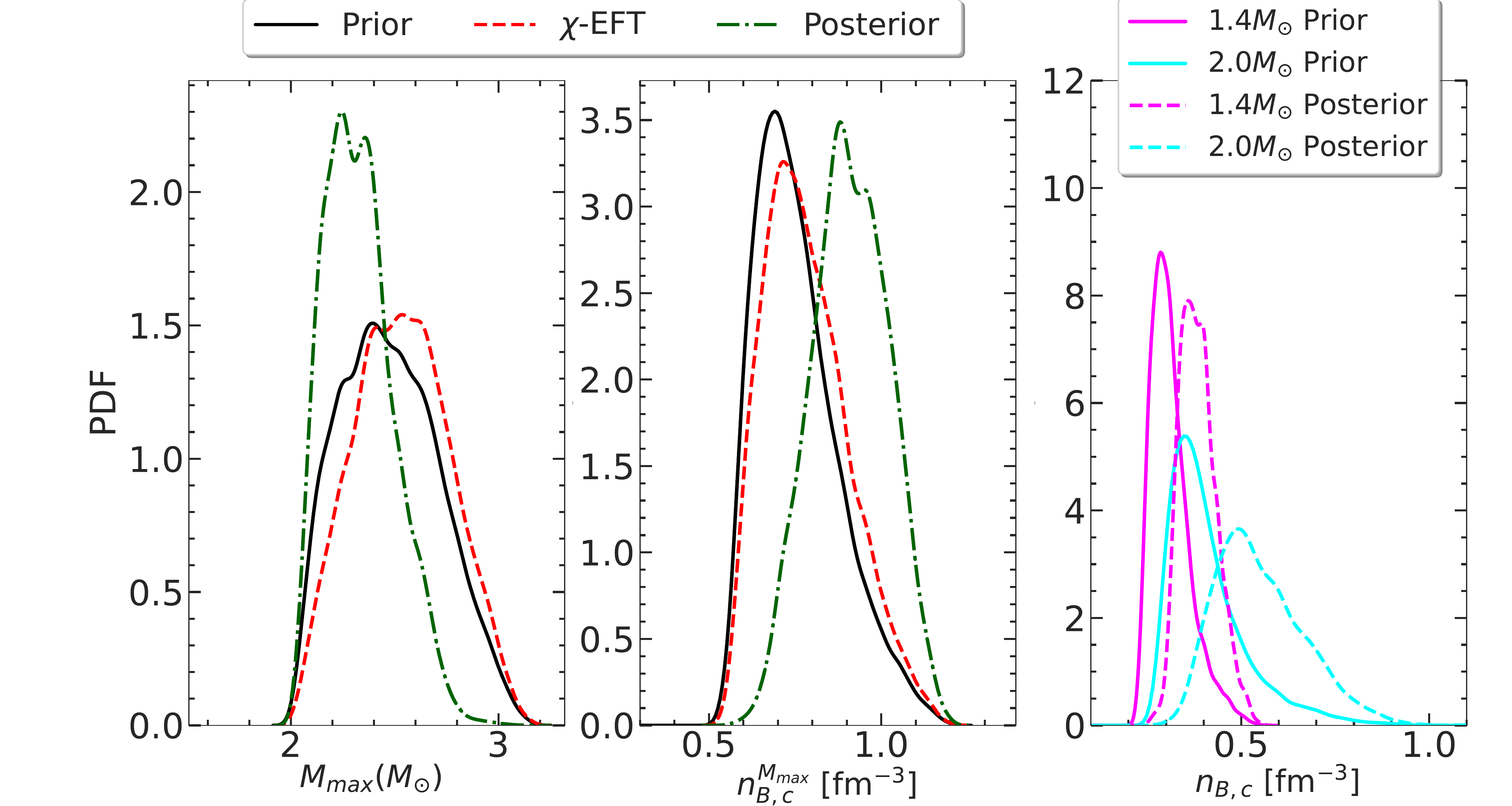}
    \caption{Distribution of maximum masses, central densities of maximum mass stars for different filters are shown on the left and middle panels, respectively. On the right panel, the distributions of the central densities of $1.4M_\odot$ and $2.0M_\odot$ stars are shown for the priors and posteriors, respectively.}
    \label{fig:mmax-central}
\end{figure}

We can also see from Fig.~\ref{fig:mrlambda} that the relativistic meta-modelling presented here can accommodate very high values of the maximum mass. Even if we have not included hyperons in the present modelling due to the high degree of uncertainty associated with their couplings~\cite{Oertel:2016bki,2015JPhG...42g5202O}, this information suggests that there will be a large parameter space accessible for the possible presence of hyperons \cite{Malik:2023mnx}.
 The marginalized distribution of the maximum mass (left), together with the associated central density (center), and the same for 1.4$M_\odot$ and 2.0$M_\odot$ NSs (right) is displayed in Fig.\ref{fig:mmax-central}. Though extreme mass values allowed in the prior are drastically reduced by the GW170817 constraint, we can see that masses compatible with the $M\approx 2.6 M_\odot$ inferred by GW190814 observation~\cite{Abbott_2020} are not excluded in our modelling. As expected, the nuclear physics knowledge embedded in the low-density filter (dashed lines in Fig.\ref{fig:mmax-central}, is not informative concerning the maximum mass. It is also interesting to observe that the central density posterior for the heaviest neutron stars peaks at extreme values $n_{B,c}^{M_{max}}\approx 0.9$ fm$^{-3}$ (middle panel), which significantly reduces to $\sim$ 0.5 fm$^{-3}$ in 2$M_\odot$ NSs (right panel). This is encouraging concerning the possibility of observing deconfined matter in the core of massive neutron stars, since the occurrence of such a phase transition is typically ruled out at low densities, even though there is a large uncertainty involved \cite{Blacker:2020nlq, Tang:2020koz,PhysRevD.103.063026}.

\subsection{Correlations between the different NMPs and NS properties}
In recent years, many efforts have been devoted to find correlations between different NMPs (see e.g.~Ref.~\cite{Mondal:2017hnh}) and between the NMPs and global NS properties, see e.g.~\cite{Malik:2018zcf,Ghosh:2021bvw}. The most prominent one is probably a correlation which has been proposed between the slope of the symmetry energy, $L_{sym}$ and the NS radius~\cite{Alam:2016cli}. It was already shown that the correlations between the NMPs observed in many nuclear interaction models is blurred by the presence of higher order NMPs, see the thourough discussion in Ref.~\cite{Margueron:2018eob}.  

\begin{figure}
    \centering
    \begin{tabular}{cc}
    \includegraphics[width=0.48\textwidth]{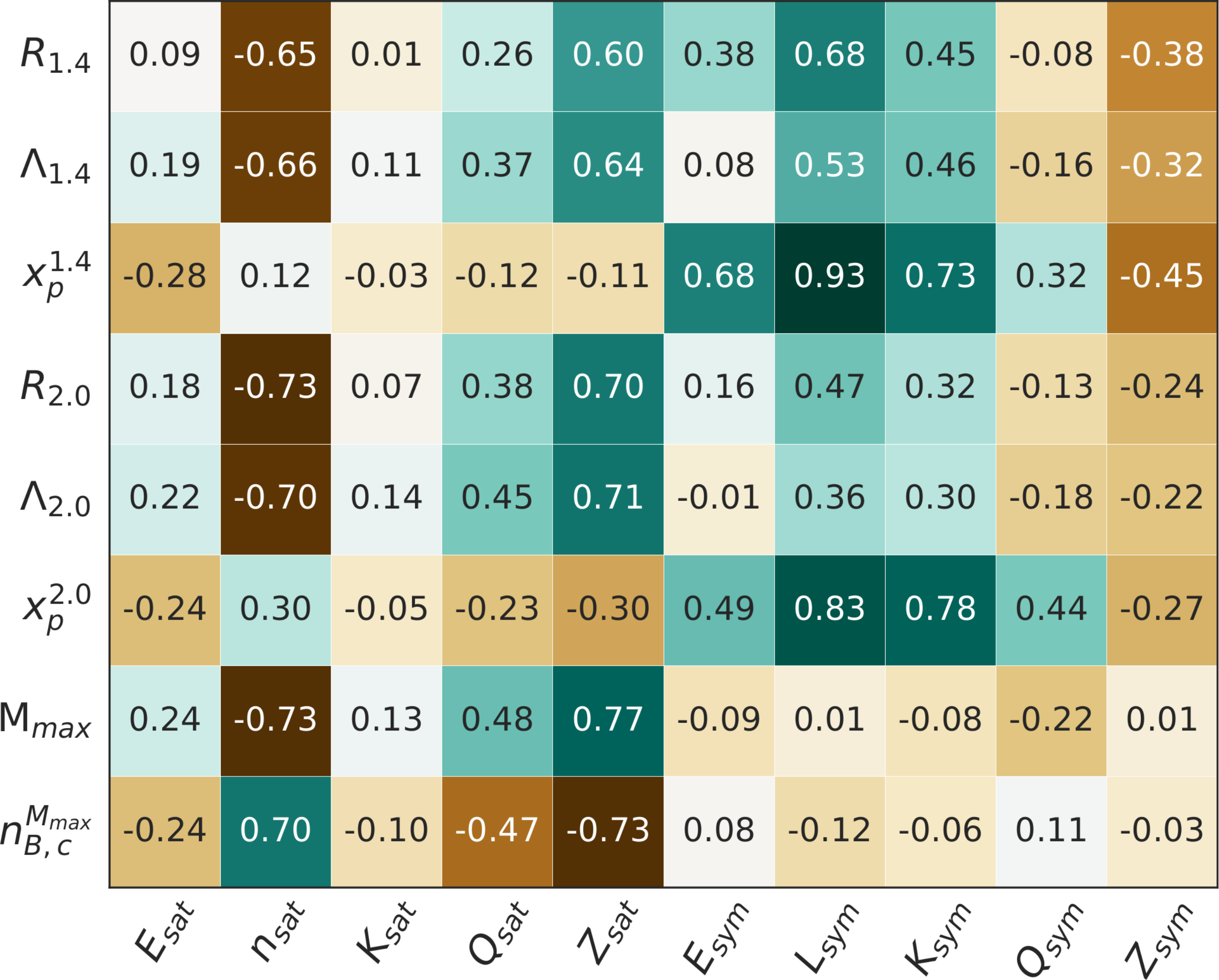}
    \includegraphics[width=0.48\textwidth]{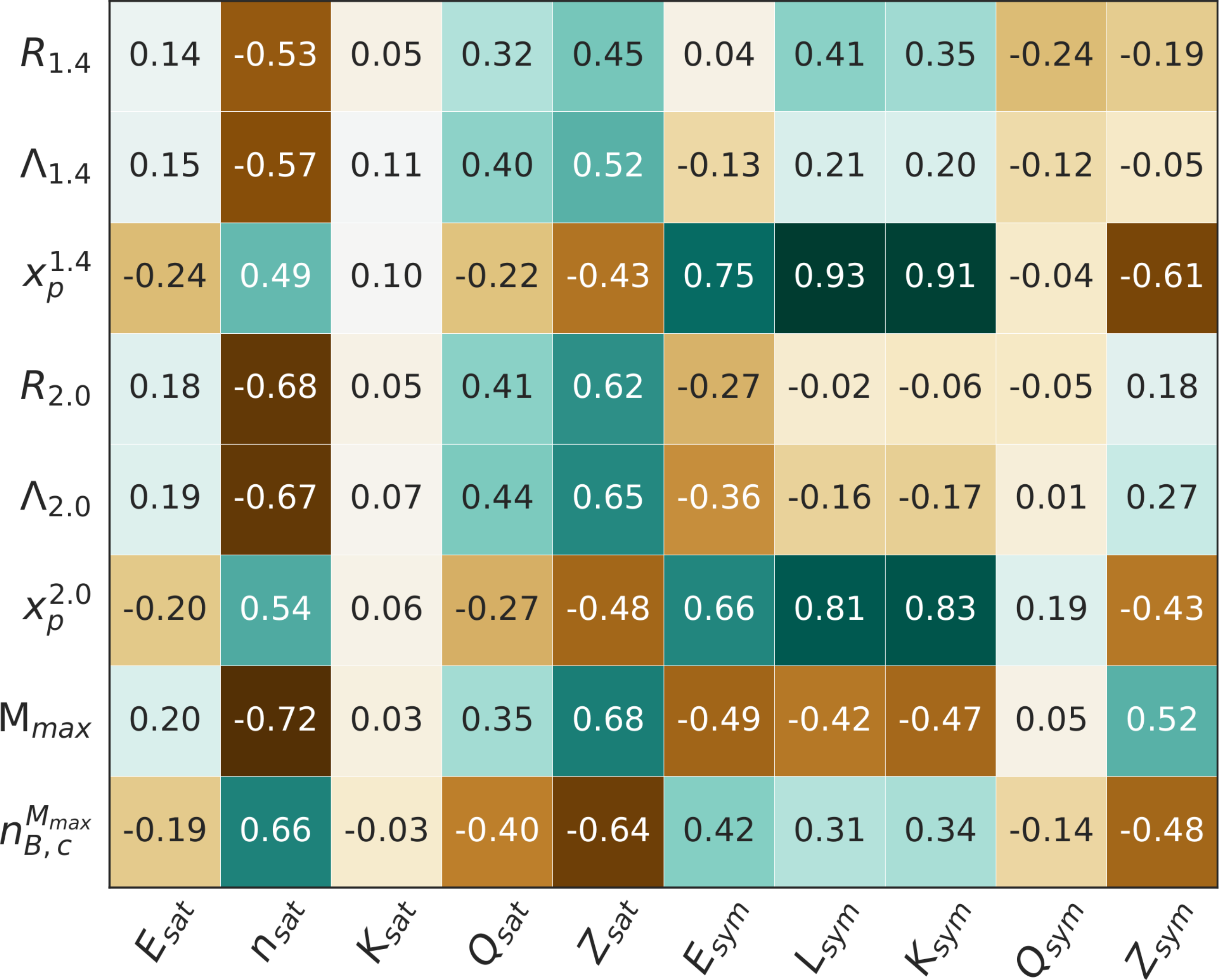} \\
    \end{tabular}
    \caption{{Pearson correlation coefficients among NMPs and some selected NS properties. (Left) Prior and (Right) Posterior.}}
    \label{fig:corr_NMP_NS}
\end{figure}

In Fig.~\ref{fig:corr_NMP_NS}, we show within the present GDFM meta-model the Pearson correlation coefficients between the NMPs and selected global NS properties such as radii, tidal deformabilities, and proton fractions for 1.4 $M_\odot$ and 2$M_\odot$ stars as well as the maximum mass and the central density of the maximum mass configuration. Again we show the results for the GDFM prior distribution and those after having applied the filters. 
{As we have already discussed, the constraints implied by the ab-initio $\chi$-EFT calculations and the low-energy nuclear experiments do not bring much information on the global star properties, see also \cite{Mondal:2022cva}. Also, the NICER measurements of neutron star radii still have too high systematics to provide effective constraints. Moreover, the maximum mass predicted by our GDFM meta-models typically exceeds the highest measured masses included in the $M_{max}$ filter Eq.(\ref{mmax}).  For this reason, the correlations observed in the final posterior as shown in the right part of the figure are essentially brought by the GW170817 tidal polarizability measurement Eq.(\ref{lvc}).}
Similar trends are observed in both cases for the majority of combinations. 
{We can see that the in-built correlations of the GDFM meta-model are typically preserved and sharpened by the observations, particularly in the isovector sector. This again suggests that they contain physical information and are not spuriously induced by the arbitrary functional form assumed for the meson couplings. The loose correlations in the prior observed in the isoscalar sector are essentially due to the nuclear mass constraints $P_{AME}$, that is already incorporated in our prior. The nuclear mass information is very effective in constraining the low-order parameters such as $n_{sat}$, $E_{sat}$ and $E_{sym}$, but this information gets diluted when the information coming from the high-order parameters is included through the astrophysical constraints.} We find very strong correlations between the central proton fractions of $1.4 M_{\odot}$ and $2 M_{\odot}$ stars with both $L_{sym}$ and $K_{sym}$ and to less extent with $Z_{sym}$. This finding is in agreement with the discussion in Sections~\ref{ssec:NMPs} and \ref{ssec:EOS} pointing out the importance of the isovector parameters for the NS interior composition.  A similar strong correlation can be seen between the central densities of the maximum mass stars and $Z_{sat}$ for both prior and posterior.  This shows the importance of this high-order NMP for the high-density part of the EOS. Interestingly, the correlation between $n^{M_{max}}_{B,c}$ and $Z_{sym}$ that is negligible in the prior, becomes substantial in the posterior. This is consistent with the strong effect of the GW170817 filter observed on the central densities in Fig.\ref{fig:mmax-central}, and underlines the effectiveness of the tidal polarizability observable to pin down the high-density behavior in the isovector sector.

\section{Conclusion}
\label{conclusion}
In this work, we have presented a relativistic generalisation of the nuclear meta-modelling technique developed in Refs.~\cite{Margueron:2017eqc, Margueron:2017lup,Carreau:2019zdy,Thi:2021jhz} for the non-relativistic case. We have discussed the importance of a relativistic approach which guarantees causality. In particular, it avoids artefacts in the sound speed distribution that could arise from excluding models which become acausal at some density in a non-relativistic setup. In a first step, we have shown that a wide exploration of the nuclear matter parameters space is possible by a judicious choice of the values for the couplings of the Lagrangian. 
For exploring the uncertainties in neutron star properties, we have imposed existing constraints on the EOSs in a Bayesian way. Then, we have explored the properties of cold infinite nuclear matter in $\beta$-equilibrium as found in mature neutron stars. Our findings indicate that the relativistic model can produce large variations of the proton fraction thus allowing for a proton-rich composition of matter at high densities in contrast to previous results which did not include enough freedom in the isovector channel. 
Filtering the models with respect to existing constraints from theoretical predictions for pure neutron matter from $\chi$-EFT, the observation of massive pulsars, and the tidal deformability of GW170817 favors lower proton fractions at high densities and reduces the variation with respect to the prior. But still relatively large proton fractions could be possible. As pointed out before by different authors, see , e.g..~\cite{2020ApJ...899....4X,Mondal:2021vzt}, to better constrain the composition of neutron star matter, additional observables sensitive to transport such as NS cooling, etc. are necessary. Concerning the NS EOS, the combined effect of the different filters of our choice select the moderately stiff EOSs from the prior GDFM models. Consequently, lower NS radii and tidal deformabilities are favored more, compared to the full space available in the prior distribution. The major impact on neutron star observables comes from the tidal deformability constraint provided by the GW170817 data, while the $\chi$-EFT filter is seen to strongly constrain the behavior of the symmetry energy.
These results are in qualitative agreement with previous findings and are compatible with NICER observations. Hence, we conclude that the GDFM meta-model can provide a robust framework to quantify uncertainties bands on NS properties. It is causal by construction and it will allow use to study exotic matters in a consistent way, which will be investigated in future. Finally, we have uploaded the EOS tables corresponding to the paramater sets, models I and II of table \ref{compare_ksym} on the publicly available \textsc{CompOSE} \cite{Typel:2013rza} database.

\section*{Acknowledgements}
This work has been partially supported by the IN2P3 Master Project NewMAC.
PC acknowledges the support of the Fonds de la Recherche Scientifique-FNRS, Belgium, under grant No. 4.4501.19. FG, CM and MO acknowledge financial support from the Agence Nationale de la recherche (ANR) under contract number ANR-22-CE31-0001-01. CM also acknowledges partial support from the Fonds de la Recherche Scientifique (FNRS, Belgium) and the Research Foundation Flanders (FWO, Belgium) under the EOS Project nr O022818F and O000422.

\bibliography{mybiblio}
\end{document}